\documentclass[11pt]{article}

\usepackage[preprint]{acl}

\usepackage{times}
\usepackage{latexsym}

\usepackage[T1]{fontenc}
\usepackage[utf8]{inputenc}

\usepackage{microtype}

\usepackage{inconsolata}

\usepackage{graphicx}
\usepackage{amsmath}        
\usepackage{booktabs}       
\usepackage{multirow}       
\usepackage{array}          
\usepackage{algorithm}      
\usepackage{algorithmic}    
\usepackage{wasysym}        
\usepackage{tcolorbox}
\usepackage{colortbl}  

\definecolor{cCtrl}{HTML}{C0C1C3}
\definecolor{cInj2}{HTML}{FDD2AA}
\definecolor{cCeleb}{HTML}{F3EFA0}
\definecolor{cInj4}{HTML}{E1B3D3}
\definecolor{cUp}{HTML}{ADCBEC}
\definecolor{dpos}{HTML}{2A6F97}    
\definecolor{dneg}{HTML}{C0504D}  

\newcommand{\cbar}[1]{\textcolor{#1}{\rule[-3.6pt]{4pt}{11.5pt}}}
\newcommand{\sw}[1]{\makebox[1.3em][l]{$^{#1}$}}
\newcommand{\nb}[1]{\makebox[2.55em][r]{#1}}

\usepackage{listings}
\tcbuselibrary{breakable,listings}
\newtcblisting{promptlisting}{
  listing only, breakable,
  colback=gray!5!white, colframe=gray!75!black,
  before upper={\ifdefined\nolinenumbers\nolinenumbers\fi},
  listing options={basicstyle=\small\ttfamily, breaklines=true, columns=fullflexible,
                   keepspaces=true, upquote=true, breakindent=0pt, breakatwhitespace=true,
                   escapeinside={(*@}{@*)}}}
\title{Using LLM Agents to Evaluate Prosocial Interventions on \\ Social Media Platforms}

\author{
  \textbf{Lucen Li}$^{1}$,
  \textbf{Erin Wertz}$^{2}$,
  \textbf{Muhammad Haroon}$^{3}$, \\
  \textbf{Anshuman Chhabra}$^{4}$,
  \textbf{Magdalena Wojcieszak}$^{2,5}$ \\
  \\
  $^{1}$Department of Computer Science, University of California, Davis \\
  $^{2}$Center for Excellence in Social Sciences, University of Warsaw \quad
  $^{3}$Meta \\
  $^{4}$Bellini College of AI, Cybersecurity, and Computing, University of South Florida \\
  $^{5}$Department of Communication, University of California, Davis \\
  \texttt{\normalsize{\{lcnli,mwojcieszak\}@ucdavis.edu}} \quad
  \texttt{\normalsize{e.wertz@uw.edu.pl}} \\
  \texttt{\normalsize{m.haroon2305@gmail.com}} \quad
  \texttt{\normalsize{anshumanc@usf.edu}}
}

\begin{document}
\maketitle
\begin{abstract}
Social media platforms are a primary source of information for many users. Although research explores various interventions to promote prosocial content or demote harmful content on platforms, there are few systematic methods for evaluating competing intervention designs before conducting costly human-subject experiments. We develop a simulation-based evaluation pipeline built on the OASIS framework. We generate 13,835 LLM-based agents, each with a unique profile. Over 20 time steps, we expose them to real-world Twitter/X posts and let them comment, like, repost, or follow. We use those agents to systematically evaluate three interventions that increase exposure to verified news: adding content, upranking content, and celebrity source cues. All interventions significantly increase news engagement, with upranking news to prominent feed positions producing the largest effects. These effects hold across agents' gender, age, and education levels. We validate our injection treatment results against a comparable human-subject study and find strong directional alignment and similar dose-normalized efficiency. Our results suggest that LLM-based agent simulations can support early-stage evaluation of platform interventions before human-subject deployment, and upranking may be more effective than news volume or source.
\end{abstract}

\section{Introduction}
Radicalization, misinformation, democratic backsliding, and the politicization of science, among others, are pressing threats worldwide. Many observers blame social media platforms, with their black-box and engagement-based algorithms, as potential culprits. Accordingly, scholars also treat these platforms as avenues for prosocial interventions to counter these threats \cite{kozyreva2024toolbox}. Interventions such as fact-checking, interface nudges, warning labels, re-ranking newsfeeds, or content replacement are hoped to reduce harmful and promote beneficial content online.

Researchers evaluate these prosocial interventions with powerful experimental methods, including online survey \citep{guess_digital_2020}, browser extension studies \cite{yu2024nudging, bhuiyanNudgeCredSupportingNews2021}, simulated platforms \cite{difranzo2018-truman}, and even---in collaboration with platforms---at-scale alteration of algorithms \cite{nyhan_like-minded_2023}. However, these studies require substantial investment in participant recruitment, infrastructure, and/or time, as well as other administrative and logistic burdens. Additionally, large scale testing may result in unintended negative effects~\cite{modirrousta2023gamified}. Agent-based modeling (ABM) offers a fast and cheap method circumventing the need for real world deployment and expensive human subjects work, but relies on highly simplified representations of users or interaction processes \cite{anwar2024filter}. 

A growing body of research finds that LLM agents can rapidly test hypotheses \cite{manning2024automated} and replicate human-experimental findings with higher verisimilitude than ABMs \cite{argyle2023out, yeykelis2024using}. At the same time, LLM agents do not behave like real people in all circumstances \cite{chuang2024simulating, zeng2026too}, so they can supplement, but not replace, human experimentation \cite{lin2025six}. 

Here, we treat LLM-based simulations as a sandbox for hypothesis generation and rapid prototyping of candidate interventions. In doing so, we extend prior literature both methodologically and substantively. \textbf{Methodologically}, we compare the efficacy of multiple interventions, within a single simulation environment, in which demographically diverse agents browse simulated newsfeeds composed of real-world Twitter/X posts rather than synthetic text. In this environment, we replicate two interventions that past studies found effective as well as a novel one. This allows us to both assess how well LLMs reproduce past field experiment results and to identify the most promising candidate for subsequent human testing.

\textbf{Substantively}, we apply this approach to increasing exposure to factual and verified public affairs information, whose consumption is low \cite{krupnikov2022other, wojcieszak2024non, thorson2021algorithmic} although it makes individuals more knowledgeable and more resistant to misinformation \cite{prior2005news, altay2024news}. We compare three interventions---direct news injection, news upranking, and celebrity source cues (news posts shown as if posted by celebrities)---against an unmodified baseline and against one another. Injection and upranking are both algorithmic manipulations \cite{yu2024nudging, askari2024incentivizing, piccardi2025reranking, stray2026prosocial}, but prior work studied them separately. Celebrity source cues parallel treatments found effective in advertising~\cite{agrawal_economic_1995} and health communication~\citep{hoffman_biological_2015} contexts but are new to news exposure. This intervention alters the source of a post rather than its visibility or the choice architecture \citep{thaler2021nudge}, allowing for a direct comparison of distinct interventions, which would traditionally be hard to study together with high verisimilitude. Although we focus on news, the mechanisms and pipeline are designed to be generalizable to other content categories and prosocial objectives.

All interventions significantly increase news engagement over an unmodified newsfeed control and the ordering holds across diverse agent demographics. Upranking produces the strongest effect, more than four times that of either four-unit news injection (Inject-4) or celebrity source cues which in turn outperform the two-unit injection condition (Inject-2). The injection effects are directionally aligned with an existing human-subject study and show similar dose-normalized efficiency. These findings show that controlled LLM environments can narrow the design space for prosocial interventions prior to expensive field deployment.

\label{sec:method}
\begin{figure*}[t]
    \centering
    \includegraphics[width=1\linewidth]{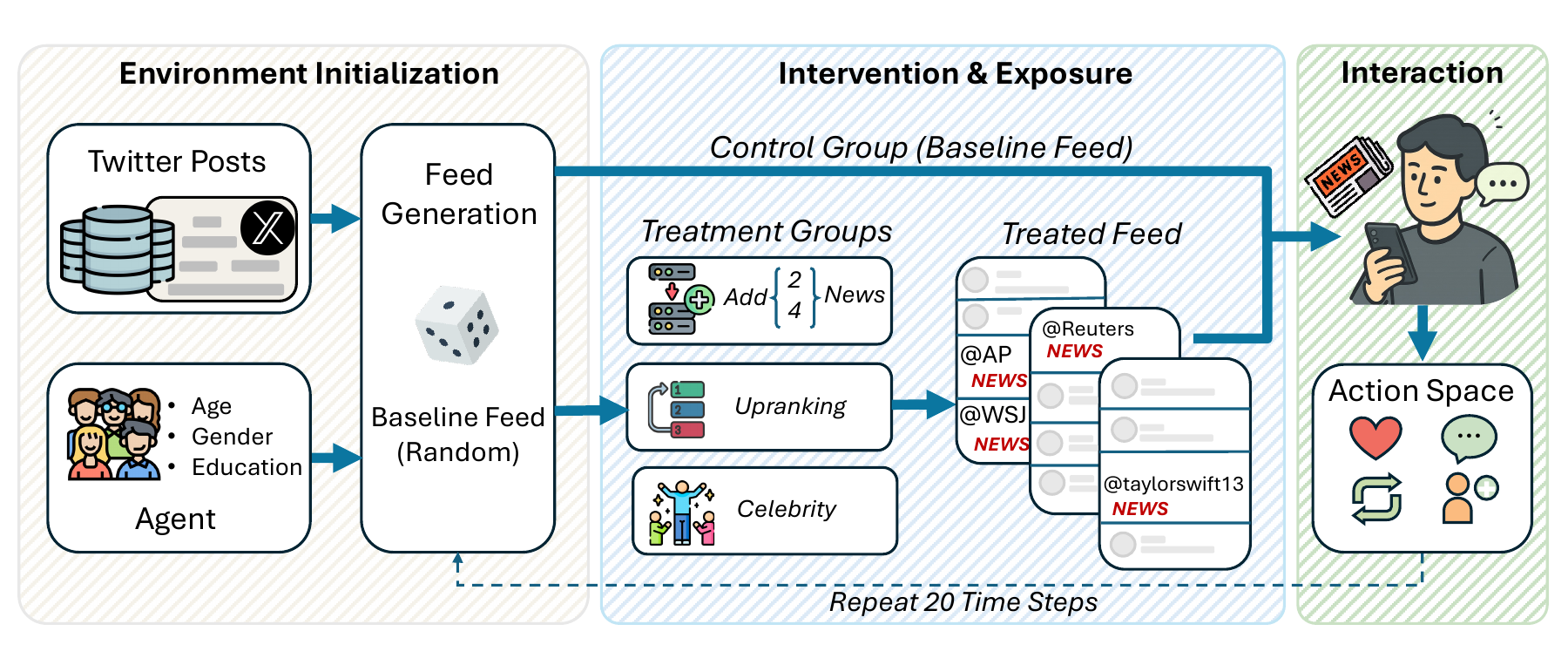}\vspace{-5mm}
    \caption{Overview of the simulation pipeline. Environment initialization stage (grey): real Twitter/X posts from the content pool, and agents are generated with age, gender, education, drawn from real demographic distributions. Intervention and exposure (blue): at each of 20 time steps, each agent receives a 40-post feed, unmodified in Control or modified by one of four interventions (Inject-2, Inject-4, Upranking, or Celebrity). Interaction (green): each agent browses the feed and selects actions from like, comment, repost or follow.}
    \label{fig:workflow}\vspace{-4mm}
\end{figure*}

\section{Related Work}
\subsection{Prosocial Interventions on Social Media}
Social media feed algorithms maximize engagement and can direct users away from high quality public affairs information toward entertainment \cite{huang2024auditing} or problematic content \cite{haroon2023auditing, haroon2022youtubegreatradicalizerauditing, habib2025youtube}.

Researchers therefore test interventions that reshape how users encounter online information \cite{kozyreva2024toolbox}. Some use browser add-ons to downrank bad or uprank good content \cite{piccardi2025reranking, munson2013encouraging}, or collaborate with platforms to test interventions such as chronological (rather than engagement-driven) feeds \cite{guess2023social} or downrank content from like-minded sources \cite{nyhan_like-minded_2023}. Others test fact-checking, digital literacy interventions, or accuracy nudges \cite{lewandowsky2021countering, arechar2023understanding, pennycook2021shifting}. 

Testing these interventions faces challenges. Browser extension studies necessitate technical skills, infrastructure, development costs, and consenting users \cite{bhuiyanNudgeCredSupportingNews2021, yu2024nudging, piccardi2025reranking}. Platform collaboration and API access are increasingly restricted \cite{mimizuka2026post}. Mock platforms, although customizable and realistic in appearance \cite{difranzo2018-truman}, struggle to recreate the diversity of engaging content. Survey experiments are easier but face external validity problems. Because each approach is expensive and time-consuming to conduct at scale, especially for researchers without substantial resources, we instead compare several candidates in a controlled environment before human testing.

\subsection{LLM-Based Agents as Social Simulators}
\looseness-1 The emergence of Large Language Models (LLMs) allows researchers to move from simple rule-based agents of traditional ABMs to ``silicon subjects'' that produce contextually rich, demographically conditioned, text-based behavior \cite{argyle2023out, aher2023using, gao2024large, mou2026individual}. Agents prompted with demographic and identifying features replicate 76\% of experimental findings in marketing research \cite{yeykelis2024using}, mirror answer distribution in large-scale surveys \cite{argyle2023out}, and model complex pandemic behavioral dynamics \cite{chopra2024limits}.

Researchers also use LLM-based social media simulations built around news content to compare platform interventions, including feed algorithms for cross-partisan dialogue \cite{tornberg2023simulating}, interventions targeting homophily and attention inequality \cite{larooij2025can}, and fact-checking strategies \cite{liu-etal-2025-mosaic}. These studies evaluate interventions primarily on outcomes such as partisan clustering, concentration of attention or the spread of misinformation; we evaluate whether engagement with the news posts themselves increases. We test three interventions that act on the volume, position, and source of those posts, a set of interventions that not previously compared in LLM-based simulation.

Despite this promise, LLM agents may generate less variance and different regression results than humans \cite{bisbee2024synthetic}; and outcomes can vary across repeated tests \cite{barrie2024replication}. Even with clear improvements compared to rule-based agents, LLMs remain ``recognizably unhuman'' \cite{zeng2026too} at best. Therefore, some suggest that LLMs can only be integrated into research if approached with a clear understanding of their limitations and strengths \cite{ma2025evaluating, lin2025six}, and that simulations should be validated for their empirical realism \cite{munker-etal-2026-dont}. Existing platform simulations validate the environment, by reproducing known social patterns \cite{larooij2025can} or matching real users' attitude trajectories \cite{mou-etal-2024-unveiling}. We focus on intervention effects, so we validate the effect of an intervention against a human field experiment. 

\section{Methodology}
\subsection{Framework}
\label{sec:framework}
We utilize the \textit{OASIS} framework \cite{yang2024oasis}, an agent-based modeling framework for social media simulation in which LLM-based agents act as users. Each agent is initialized with its LLM-generated real name, user name, and demographic attributes (age, gender, education). We run the simulation over 20 discrete time steps. At each time step, each agent receives a feed randomly generated by the system, and processes the entire feed in a single inference call. Based on its demographic attributes, the full feed content, and a memory window containing the feeds encountered and actions previously taken, the agent performs cognitive reasoning and generates all engagement decisions at once, selecting which posts to interact with and which actions to perform (like, comment, repost, follow). The agent's state is then updated for subsequent steps. Figure \ref{fig:workflow} provides an overview of the simulation pipeline. Additional implementation details for each step are provided in Appendix \ref{app:oasis}. 

\subsection{Agent Design}
 We generate 15,000 LLM agents, of which 13,835 are retained for the final analysis; we exclude agents that took no action at any of the 20 time steps. Each agent is assigned an age, gender and education sampled from a real-world distribution (see Appendix \ref{app:agents} for details). These attributes are then used as prompts for GPT-4o-mini (temperature of 0.8) to synthesize realistic usernames and display names (prompt in Appendix \ref{app:profile}). 
 
 During the simulation, all agents run on GPT-4o-mini (temperature of 0.7) and are instructed to act naturally as Twitter/X users: at each time step, each agent browses a newly generated newsfeed (as described below) and reports whether they \textit{like, comment, repost} any post in the feed, or \textit{follow} post authors. Agents retain memory of up to the most recent five time-steps. Appendix \ref{app:timeline} provides the details of the memory manipulation. 

\subsection{Simulation Environment}
\paragraph{Content Pool.} We construct the content pool from two sources. First, we collect 477,480 English-language tweets from Twitter/X using Apify\footnote{\url{https://apify.com/apidojo/tweet-scraper}}. We exclude reposts, replies, quote tweets that depend on conversation context, as well as posts containing external links or images unavailable in the text-only prompts. This yields 435,273 tweets from 12,068 accounts. Second, to provide a pool of verified news posts for the feeds and interventions, we supplement these with 32,039 tweets from 35 news accounts with high quality ratings from \citet{lin2023high} and low bias ratings \cite{adfontes2025}, drawn from a publicly available corpus of tweets\footnote{\url{https://huggingface.co/datasets/enryu43/twitter100m_tweets}}. The combined pool contains 467,312 tweets from 12,103 accounts. Appendix \ref{app:news_list} lists the outlet and selection details. 

\paragraph{Content Labeling.} We label all posts from the 35 outlets as news and political, and label the remaining tweets as political by a RoBERTa-based classifier \cite{askari2024incentivizing}. Appendix \ref{app:pol_class} provides details about the classifier and example of the labeling. 

\paragraph{Feed Construction.} At each time step, we sample a 40-tweet feed for each agent, with an expected political share of 10\%, approximating to the 8\% share of political tweets found in users' Twitter/X timelines \cite{pew2022}. Each tweet is assigned a random timestamp between May 1 and May 3 2023 so that posts in a feed are temporally proximal to one another, as in real newsfeeds; the feed is ordered from newest to oldest. Each tweet shows its author, text, timestamp and counts of likes, reposts and comments. Appendix \ref{app:feed} details the sampling procedure and timestamps.
 
\begin{figure*}[t]
  \centering
  \includegraphics[width=0.8\textwidth]{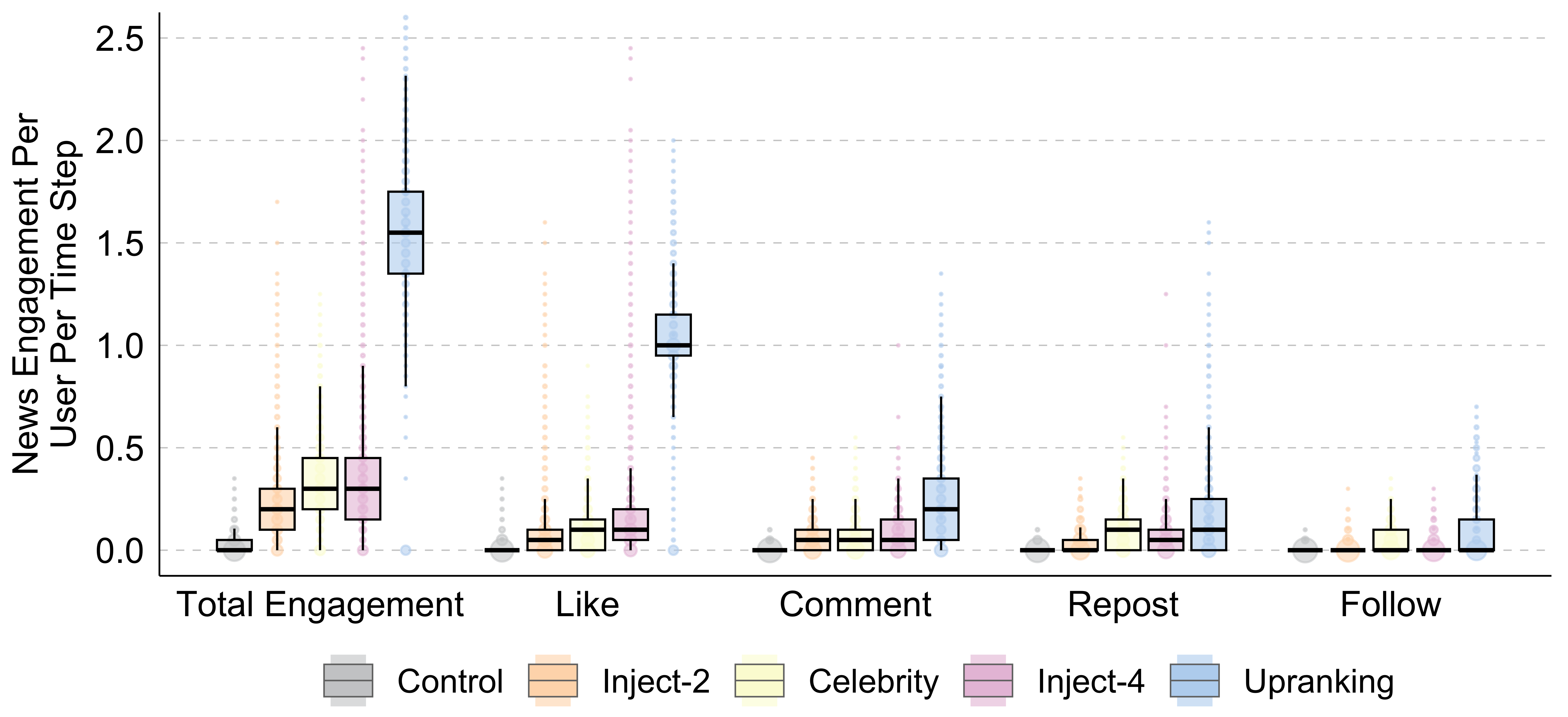}\vspace{-4mm}
  \caption{Distribution of raw news engagement counts by treatments for five engagement metrics (Total Engagement, Likes, Comments, Reposts, and Follows). Horizontal lines mark medians; each semi-transparent dot is one agent's engagement count at one time step. Control agents (grey) rarely engage with news; Upranking (blue) exhibits the largest shift.}
  \label{fig:raw_box}\vspace{-3mm}
\end{figure*}
\subsection{Treatment Design}
Agents are randomly assigned to one of five conditions (Control, Celebrity, Inject-2, Inject-4, or Upranking) and remain in it for all 20 time steps. Final conditions contain 2,447, 2,831, 2,852, 2,827 and 2,878 agents, respectively. %Appendix \ref{app:treatment} provides the implementation details of each treatment. 
The control condition shows the unmodified feed as described above; while each treatment condition modifies the agent's own feed as described next. To ensure comparability across treatment conditions, all treatment groups share a common injection manipulation, operated by a \textit{Shared News Payload}. At each time step, the system samples four news posts $\mathcal{N}^{(4)}_t \subset \mathcal{P}_{\text{news}}$. These are injected into the recommendation sequence through an operation $\Phi(\mathcal{R}, m)$ that replaces $m$ randomly selected posts while keeping the feed length fixed:
\begin{equation*}
\Phi(\mathcal{R}, m) \rightarrow \mathcal{F} \quad \text{s.t.} \quad \mathcal{N}^{(m)} \subset \mathcal{F}, \; |\mathcal{F}| = |\mathcal{R}|
\end{equation*}
The first two posts from $\mathcal{N}^{(4)}_t$ are shared across Inject-2, Upranking, and Celebrity; all four are used by Inject-4. Inject-2 therefore serves as the \textit{Reference Treatment}. Comparing each of the other treatments with Inject-2 isolates the effect of its specific modification. We denote the resulting visible \textit{feed} as $\mathcal{F}_{i,t}$, distinguishing it from the pre-treatment \textit{recommendation sequence} $\mathcal{R}_{i,t}$.

\paragraph{Inject News Interventions.} Injection manipulates the \textit{volume} of news, which field experiments find to increase news exposure and engagement \cite{yu2024nudging, askari2024incentivizing}. The injection conditions show:
\begin{equation*}\small
\mathcal{F}_{i,t}^{\text{Inject-m}} = \Phi(\mathcal{R}_{i,t}, m), m\in \{2,4\}
\end{equation*}
So the comparison between the Inject-2 and Inject-4 isolates the marginal effect of doubling the injected news volume.

\paragraph{Upranking Intervention.} Upranking manipulates the \textit{positional visibility} of news. Posts at the top of feeds attract substantively more engagement \cite{chan2025ranking} and prior studies found reranking to be a successful intervention for other types of content \cite{piccardi2025reranking, stray2026prosocial}. The treatment applies $\Phi(\mathcal{R}, 2)$, as in Inject-2, and then moves the two injected news posts to the most prominent feed positions (ranks 0 and 2):
\begin{equation*}\small
\text{\small$\displaystyle \mathcal{F}_{i,t}^{\text{Uprank}} = \Phi(\mathcal{R}_{i,t}, 2)\;\;\text{s.t.}\;\; \mathrm{rank}(n) \in \{0,2\}\ \forall n \in \mathcal{N}^{(2)}$},
\end{equation*}
where $rank(n)$ is the position of post $n$ in the feed; a post already at its target position stays in place.

\paragraph{Celebrity Source Cue Intervention.}
The celebrity source cues manipulate the \textit{source} of news. Celebrity endorsements attract attention and are widely used in health and environmental campaigns \cite{olmedo2020scoping, hoffman_biological_2015}, but are untested for news. The treatment applies $\Phi(\mathcal{R}, 2)$, as in Inject-2, and replaces only the displayed source of each injected news posts:
\begin{equation*}\small
\mathcal{F}_{i,t}^{\text{Celeb}} = \tau(\Phi(\mathcal{R}_{i,t}, 2), \mathcal{N}^{(2)}),
\end{equation*}
where $\tau$ replaces the displayed author name and username of posts in $\mathcal{N}^{(2)}$ with celebrity identities with a verified label selected from the ten most-followed English-language accounts on Twitter/X, balanced by gender (Appendix~\ref{app:celebrity}). Because content, quantity, and position match Inject-2, differences in engagement are attributable to the source cue, including its verified label. 

\subsection{Statistical Analysis}
\label{sec:stats}
\paragraph{Model Specifications and Treatment Effects.} Our primary outcome is the \textit{count} of news engagement actions (like, comment, repost, or follow on a news account) by each agent at each time step. Because this outcome is a non-negative integer with a right-skewed distribution, we model it using a Poisson generalized linear mixed model (GLMM) \cite{brooks2017glmmtmb}:
\begin{equation}
\begin{split}
\log(\lambda_{it}) =
&\ \beta_0 + \beta_1 \text{Treat}_i + \beta_2 \text{Age}_i \\
&+ \beta_3 \text{Gender}_i + \beta_4 \text{Edu}_i + u_i .
\end{split}
\label{eq:poisson}
\end{equation}
where $Y_{it}$ is agent $i$'s news engagement count at time step $t$, $\text{Treat}_i$ is the intervention condition, $\text{Age}$, $\text{Gender}$ and $\text{Edu}$ are demographic covariates, and $u_i \sim \mathcal{N}(0, \sigma^2)$ is a per-agent random intercept (20 observations per agent). To evaluate whether and how agents shift their engagement with news within their overall activity, we also estimate a \textit{rate} model that adds an offset term $\log(T_{it})$, where $T_{it}$ is the agent $i$'s total action count at step $t$. For both count and rate specifications, we report estimated marginal means (EMMs), the model-predicted expected news engagement per step (count model) or share of news actions (rate model) for each treatment, averaged over the demographic covariates. EMMs are computed with the emmeans package \cite{lenth2025emmeans}, with age as its mean, demographic categories weighted equally, and the random intercept set to zero. Pairwise contrasts between conditions are computed on the response scale with Tukey-adjusted $p$-value to control the family-wise error rate across multiple comparisons. Robustness checks using alternative model specifications are reported in Appendix~\ref{app:full_stats}.

\paragraph{Benchmarking Strategy. } We benchmark the injection conditions (Inject-2 and Inject-4) against human-subject results by \citet{yu2024nudging}, because their intervention, like ours, increased the share of news in users' feeds. The two studies measure different outcomes (news watching in \citet{yu2024nudging} and engagement actions in ours). We therefore treat both as indicators of attention to news and compare each study's treatment effect to its own control baseline, on both count and rate scales. Finally, to account for varying intervention intensities (e.g., injection dosage), we compute a dose-normalized efficiency for each study: the percentage-point (pp) increase in news engagement share divided by the pp increase in news exposure share between treatment and control groups. Appendix \ref{app:validation} details the estimation procedures, model specifications, and robustness check with alternative models.

\section{Results}
\begin{table*}[t]
  \centering
  \footnotesize
  \setlength{\tabcolsep}{5.5pt}
  \renewcommand{\arraystretch}{1.04}
  \caption{Poisson GLMM coefficients are on the log scale (standard errors in parentheses) from five models. Intervention effects are relative to Control (Pred./step$=0.027$); demographic effects are relative to female and to a bachelor's degree. Pred./step is the model-implied news engagement per agent per step (EMM) from the Total model. Demographic cells are colored by the coefficient's magnitude: blue (positive) or red (negative), darker for larger $|\beta|$. \textbf{Bold} and \underline{underline} mark the largest and second-largest treatment effect per column. All treatments significantly increase news engagement relative to Control. *** $p < 0.001$, ** $p < 0.01$, * $p < 0.05$.}
\label{tab:main_results}
  \begin{tabular}{@{}c@{\hspace{4pt}}l@{\hspace{6pt}}l l l l l l@{}}
    \toprule
    & & Pred./step & \multicolumn{1}{c}{Total} & \multicolumn{1}{c}{Likes} & \multicolumn{1}{c}{Comments} & \multicolumn{1}{c}{Reposts} & \multicolumn{1}{c}{Follows} \\
    \midrule
    \cbar{cInj2} & Inject-2 & 0.227 & \nb{2.136}\,\sw{***}(.030) & \nb{1.957}\,\sw{***}(.043) & \nb{2.755}\,\sw{***}(.076) & \nb{2.099}\,\sw{***}(.067) & \nb{1.350}\,\sw{***}(.100) \\
    \cbar{cCeleb} & Celebrity & 0.347 & \nb{2.560}\,\sw{***}(.030) & \nb{2.022}\,\sw{***}(.043) & \nb{2.893}\,\sw{***}(.075) & \nb{\underline{3.085}}\,\sw{***}(.065) & \nb{\underline{2.962}}\,\sw{***}(.094) \\
    \cbar{cInj4} & Inject-4 & 0.363 & \nb{\underline{2.605}}\,\sw{***}(.029) & \nb{\underline{2.453}}\,\sw{***}(.042) & \nb{\underline{3.115}}\,\sw{***}(.075) & \nb{2.637}\,\sw{***}(.066) & \nb{1.886}\,\sw{***}(.097) \\
    \cbar{cUp} & Upranking & \textbf{1.618} & \nb{\textbf{4.100}}\,\sw{***}(.029) & \nb{\textbf{4.395}}\,\sw{***}(.041) & \nb{\textbf{4.130}}\,\sw{***}(.074) & \nb{\textbf{3.425}}\,\sw{***}(.065) & \nb{\textbf{3.365}}\,\sw{***}(.093) \\
    \midrule
    & \multicolumn{2}{@{}l}{Age: Centered} & \cellcolor{dpos!6}\nb{.002}\,\sw{***}(.000) & \cellcolor{dpos!6}\nb{.002}\,\sw{***}(.001) & \cellcolor{dpos!6}\nb{.005}\,\sw{***}(.001) & \cellcolor{dpos!6}\nb{.003}\,\sw{***}(.001) & \cellcolor{dneg!7}\nb{$-$.011}\,\sw{***}(.002) \\
    \addlinespace[2.5pt]
    & \multicolumn{2}{@{}l}{Gender: Male} & \cellcolor{dneg!11}\nb{$-$.054}\,\sw{***}(.009) & \cellcolor{dneg!10}\nb{$-$.048}\,\sw{**}(.015) & \cellcolor{dneg!13}\nb{$-$.081}\,\sw{***}(.019) & \cellcolor{dneg!14}\nb{$-$.090}\,\sw{***}(.019) & \nb{.000}\,\sw{}(.038) \\
    \addlinespace[2.5pt]
    & \multicolumn{2}{@{}l}{Edu: Doctoral} & \cellcolor{dpos!13}\nb{.076}\,\sw{***}(.017) & \cellcolor{dpos!13}\nb{.078}\,\sw{**}(.028) & \cellcolor{dpos!19}\nb{.143}\,\sw{***}(.034) & \cellcolor{dpos!9}\nb{.031}\,\sw{}(.035) & \cellcolor{dneg!23}\nb{$-$.187}\,\sw{*}(.074) \\
    & \multicolumn{2}{@{}l}{Edu: Master's} & \cellcolor{dpos!11}\nb{.051}\,\sw{***}(.014) & \cellcolor{dpos!11}\nb{.049}\,\sw{*}(.024) & \cellcolor{dpos!14}\nb{.092}\,\sw{**}(.029) & \cellcolor{dneg!9}\nb{$-$.032}\,\sw{}(.031) & \cellcolor{dpos!11}\nb{.053}\,\sw{}(.061) \\
    & \multicolumn{2}{@{}l}{Edu: Some college} & \cellcolor{dneg!16}\nb{$-$.103}\,\sw{***}(.013) & \cellcolor{dneg!15}\nb{$-$.100}\,\sw{***}(.022) & \cellcolor{dneg!25}\nb{$-$.205}\,\sw{***}(.028) & \cellcolor{dneg!20}\nb{$-$.149}\,\sw{***}(.028) & \cellcolor{dpos!20}\nb{.156}\,\sw{**}(.055) \\
    & \multicolumn{2}{@{}l}{Edu: Some high school} & \cellcolor{dneg!16}\nb{$-$.112}\,\sw{***}(.013) & \cellcolor{dneg!12}\nb{$-$.063}\,\sw{**}(.022) & \cellcolor{dneg!32}\nb{$-$.287}\,\sw{***}(.028) & \cellcolor{dneg!26}\nb{$-$.219}\,\sw{***}(.029) & \cellcolor{dpos!15}\nb{.102}\,\sw{}(.055) \\
    & \multicolumn{2}{@{}l}{Edu: No formal education} & \cellcolor{dneg!14}\nb{$-$.089}\,\sw{***}(.016) & \cellcolor{dpos!6}\nb{.005}\,\sw{}(.026) & \cellcolor{dneg!36}\nb{$-$.325}\,\sw{***}(.034) & \cellcolor{dneg!30}\nb{$-$.263}\,\sw{***}(.035) & \cellcolor{dpos!12}\nb{.064}\,\sw{}(.066) \\
    \bottomrule
  \end{tabular}
\end{table*}
%\begin{table*}[t]                                             
\subsection{Treatment Effects on News Engagement}
\label{sec:overall}
We first examine whether each treatment increases agents' engagement with news. Figure \ref{fig:raw_box} shows the raw engagement counts. Most agents in the control group exhibit minimal interaction with news, with the majority of agents remaining inactive, and every treatment increases engagement counts across all action types. 

\paragraph{Overall Treatment Effects.} We estimate the effect of each intervention on \textit{per-step aggregated news engagement}, the number of news engagements (like, comment, repost, follow) an agent takes at each time step, using the Poisson mixed-effects model described in Section \ref{sec:stats}. News engagement in the control condition is extremely sparse, with an EMM of 0.027 news engagements per step. As shown in Table \ref{tab:main_results}, all four interventions lead to positive and statistically significant increase in per-step news engagement. For example, Inject-2 treatment ($\beta=2.136$, $SE=0.030$, $p<0.001$) raises predicted news engagement to approximately 0.227 actions per step, while the Celebrity treatment ($\beta=2.560$, $SE=0.030$, $p<0.001$) and Inject-4 ($\beta=2.605$, $SE=0.029$, $p<0.001$) both raise engagement to approximately 0.347 and 0.363 actions per step, respectively. Upranking ($\beta=4.100$, $SE=0.029$, $p<0.001$) yields the largest effects, increasing predicted engagement to 1.618 actions every step. We also examine the rate of news engagement, defined as news engagements as a share of all actions per agent per step, using total engagement as an offset. The predicted share of news engagements rises from 1.1\% in the control condition to 7.6\% under Inject-2, 12.0\% under Inject-4, 15.0\% under Celebrity, and 40.8\% under Upranking, with treatment effects remaining positive and statistically significant across all conditions. Full results are provided in Table \ref{tab:rate_poisson}. All findings are robust to analytical approach with alternative models showing consistent effects of both the treatments and agents' demographics (see Appendix \ref{app:full_stats}).

\paragraph{Demographic Effects.} In the Total-engagement model, all demographic predictors including age, gender and education are statistically significant ($p<0.001$) and broadly consistent with demographic effects on news consumption identified in past literature, where older and more educated individuals tend to consume more news \cite{boulianne2022age}. However, the magnitude of these effects is much smaller than that of our treatments. Even the largest difference observed, i.e., those with doctoral degrees (.076 vs bachelor's) and agents with some high school (-.112 vs bachelor's), is .188 log-scale engagements, is much smaller than even the smallest treatment effect observed ($\beta=2.136$, Inject-2). We further test for potential heterogeneity in treatment effects across demographic subgroups, but find no statistically significant evidence of interaction effects (all $|\beta| < 0.15$; See Appendix~\ref{app:demographic}). 

As a robustness check, we replicate all analyses using engagement with all political posts in the feed, not only with news posts, finding comparable effects. This ensures that the effects we observe are not driven by the low baseline frequency of news posts in sample feeds prior to treatments (see Appendix \ref{app:robust_political_engagement} for details).

\subsection{Relative Effectiveness of Three Treatment Mechanisms}
\label{sec:relative}
We next compare the treatments with one another. Because every treatment injected at least two news posts, contrasts against Inject-2 isolate the effect of each treatment's own manipulation: doubled volume, position or source cue. Contrasts among the other treatments investigate which treatment(s) are \textit{most} effective relative to all others.  

\begin{figure}[t]
  \centering
  \includegraphics[width=\columnwidth]{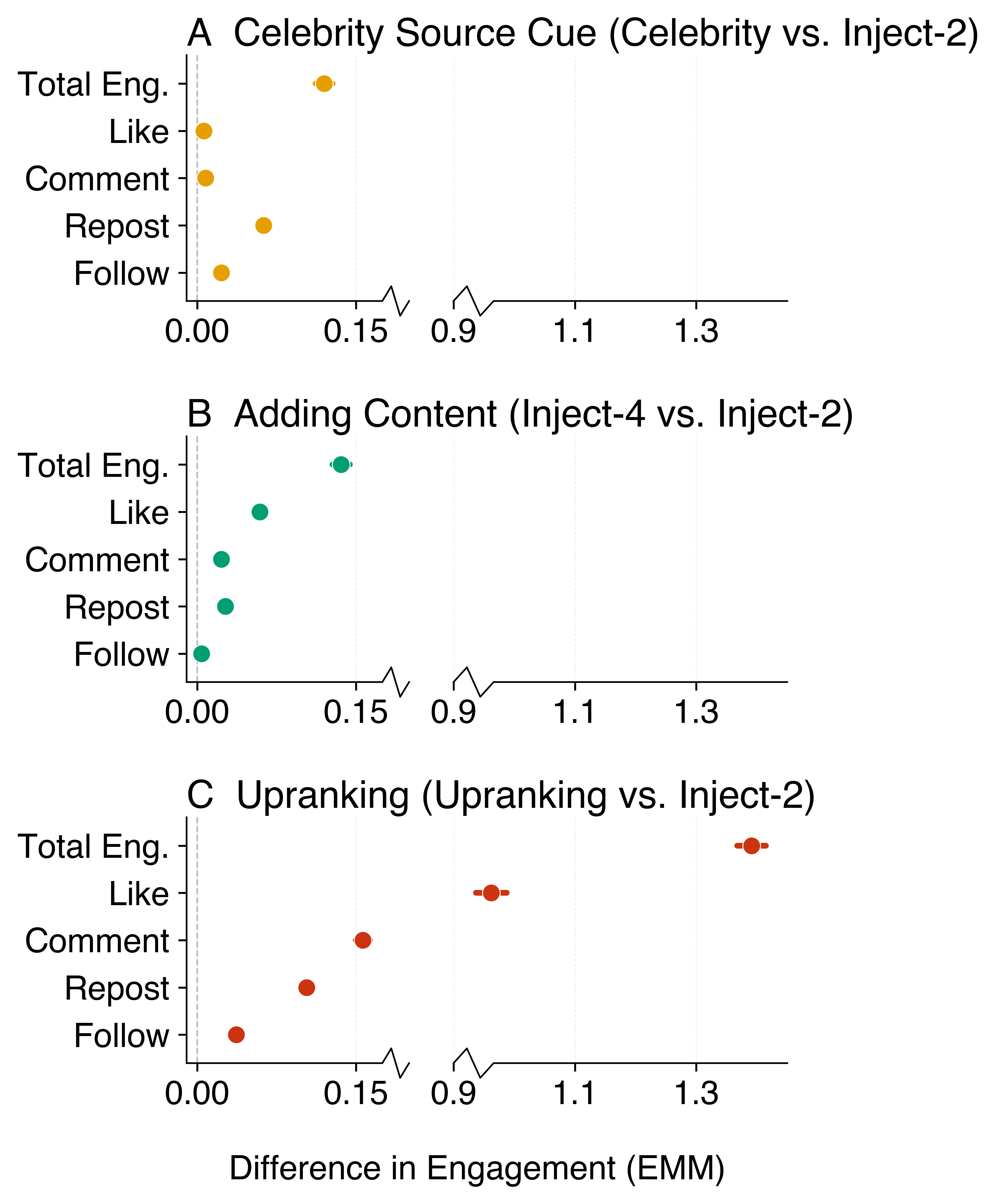}\vspace{-3mm}
  \caption{Post-hoc Pairwise Contrasts of EMMs by Engagement Types. Panels display the contrast estimates between treatment conditions for total engagement, like, comment, repost, and follow counts. Points represent the estimated differences, with error bar indicating 95\% CI. The vertical dashed line at zero represents the null hypothesis. P-values and CIs are Tukey-adjusted.}
  \label{fig:pairwise}\vspace{-3mm}
\end{figure}

\begin{figure}[t]
  \centering
  \includegraphics[width=0.85\columnwidth]{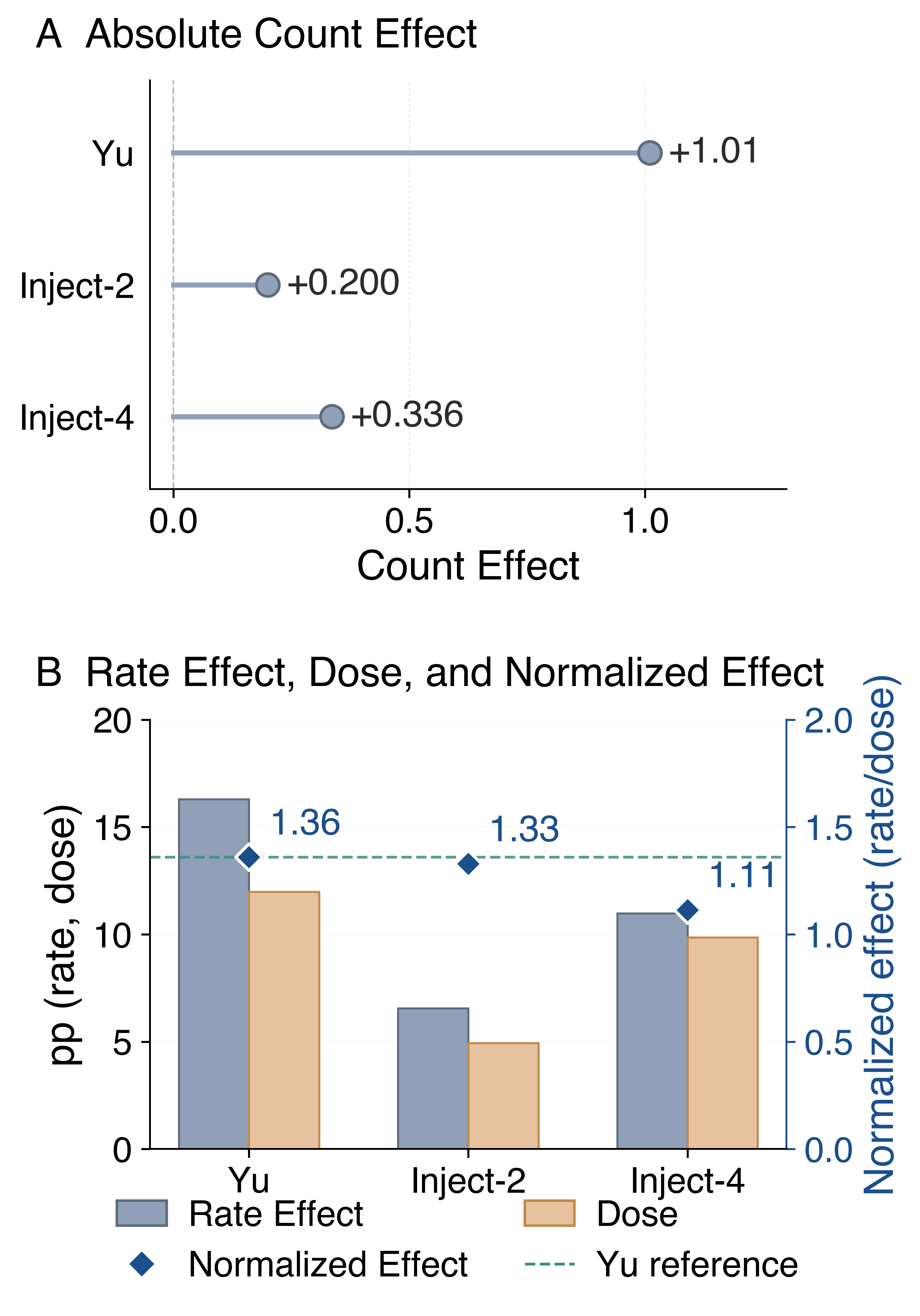}\vspace{-3mm}
  \looseness-1\caption{Validating simulation results against \citet{yu2024nudging}. Panel A presents count-based treatment effects. All three conditions show significant positive effects, confirming directional consistency. Panel B presents dose-normalized rate-based comparison. The $x$-axis shows the between-group increase in news feed share (dose); the $y$-axis shows the share-based treatment effect. All treatments are statistically significant ($p < 0.001$).}
  \label{fig:yu}\vspace{-4mm}
\end{figure}

Figure~\ref{fig:pairwise} shows the EMM comparisons against Inject-2 condition. All three treatment mechanisms produce statistically significant effects relative to Inject-2 (all $p < 0.001$).

\looseness-1 On total engagement count, the celebrity source cue produces an additional $0.120$ ($SE=0.004$, $p < 0.001$) engagements per step beyond the Inject-2 baseline, indicating that displaying news content as coming from a celebrity meaningfully increases engagement. Increasing the exposure frequency (doubling the number of injected news from 2 to 4) leads to an incremental gain of $0.136$ ($SE=0.004$, $p < 0.001$) total reactions compared to the Inject-2 baseline. Upranking news content in a feed generates $1.391$ ($SE=0.012$, $p<0.001$) more engagements per time step than the Inject-2 baseline. 

\looseness-1 We additionally compare all treatments to one another. For total engagement, the absolute difference between Inject-4 and Celebrity is substantively negligible ($0.016$); although it reaches statistical significance due to our large sample size, the two mechanisms are comparable in their effects. In contrast, upranking consistently outperforms both Inject-4 and Celebrity interventions, generating an additional $1.255$ ($SE=0.012$, $p < 0.001$) and $1.271$ ($SE=0.012$, $p < 0.001$) news engagements per step, respectively. Agents' responsiveness to high-ranking content underscores that the visibility of news in the feed matters far more than its quantity or source. Action-specific EMMs on both count and rate scales are reported in Appendix \ref{app:action_specific}.

\subsection{Benchmarking Against Human-Subject Findings}
\label{sec:validation}
We validate our injection conditions against \citet{yu2024nudging}'s finding that nudging the recommendation algorithm to show more news increased news consumption on YouTube. This allows us to investigate whether the simulated effects are directionally consistent and comparable in relative magnitude.

On a count basis, \citet{yu2024nudging}'s treatment group watched +1.01 more news videos per active day than control ($p < 0.001$), and our injection conditions add +0.200 (Inject-2) and +0.336 (Inject-4) news engagements per step (both $p < 0.001$). Both results show positive and significant effects, but the absolute counts are not directly comparable, because a YouTube active day involves substantially more content than a 40-post step. On a rate basis, \citet{yu2024nudging}'s treatment raised the proportion of news in total video consumption by +16.29 pp ($p < 0.001$), and our conditions raise the share of news in engagement by 6.55 pp (Inject-2) and 10.98 pp (Inject-4). 

To compare the effects on a common scale, we conduct a dose-normalized comparison: divide each effect by its dose. The between-group increase in share of news in the feed is $+11.99$ pp for \citet{yu2024nudging}, computed from their public data; in our simulation, the effects and are $+4.93$ pp (Inject-2) and $+9.86$ pp (Inject-4). As shown in Figure~\ref{fig:yu}, both count-based and rate-based treatment effects are directionally consistent and statistically significant across the simulation and the field experiment. After normalizing for intervention dose, the resulting per-unit efficiency is closely aligned: 1.33 (Inject-2) and 1.11 (Inject-4) vs. 1.36 in the field experiment. 

\section{Discussion}
\label{sec:discussion}
Upranking produces the largest effect in our simulation: agents engage with news 4.46 and 4.67 times often as under Inject-4 and Celebrity, respectively, because agents' engagement is concentrated at the top of the feed. In the control condition, 70\% of engagements fall on the first three positions, and a post at position 0 receives the same engagement whether or not it is news or not. Appendix \ref{app:position} provides the position curves. Human users show the same direction of position effect \cite{chan2025ranking}, but the mechanism differs: humans scroll scan a feed, whereas LLM agents process the feed as an input and shows position effects for long inputs \cite{liu2024lost}. Thus, the absolute magnitude of the upranking effect should be interpreted with caution; separating it would requires simulations that decouple feed rank from prompt order, followed by human validations.

The celebrity source cue intervention modestly increase engagement with the same news attributed to news organizations, with effects concentrated in reposts and follows rather than likes or comments. Because following is a one-time action, this effect reflects agent's initial response more than sustained attention to the news itself. As no directly comparable human study exists, this result functions as supporting evidence that this particular treatment is likely to bear fruit in subsequent testing, rather than confirmative evidence that celebrities posting news stories increase engagement. Finding similar effects would strengthen the case for the simulation's predictive validity, or identify a case where agent and human behavior diverge.

The injection effects align with \citet{yu2024nudging} in direction and, after dose normalization, in efficiency, but not in raw effect size, as in other LLM-agent replications \cite{ma2025evaluating}. The findings should be interpreted in light of several design choices inherent to the LLM agent-based simulation approach. A strong future test of the applicability of LLM agents for intervention testing would be to use the LLM-based agent simulation to generate predictions first and then run a human subject experiment to check whether those predictions hold among actual social media users.

\section{Conclusion}
\label{sec:conclusion}
Methodologically, we present an LLM agent-based simulation framework for evaluating prosocial interventions on social media, applied to increasing user engagement with verified public affairs information. The framework compares multiple interventions without recruiting human participants, and can rank candidate interventions within one environment, before costly field deployment. The injection effects align with a human-subject field experiment in direction, and dose-normalized efficiency, supporting LLM-based simulation as a piloting tool for intervention design. Substantively, our simulation shows that upranking news to prominent positions is substantially more effective than increasing the volume of news content or attaching celebrity source cues to identical information. It also generates an untested prediction that celebrity source cues modestly increase news engagement, offering a target for future human testing. The pipeline's modular design allows the target content category and intervention mechanisms to be reconfigured for broader challenges of content promotion on social media.

\section*{Limitations}
First, agents are characterized by minimal demographic attributes (age, gender, education), and their action space is limited to four engagement types (like, comment, repost, follow), restricting the behavioral diversity of the simulated population. Feed construction and treatment implementation lack the personalization and feedback loops of real social media platforms. Because all conditions share the same feed generation process, these simplifications affect absolute engagement effects more than the differences between conditions, which are our main findings.

Second, our findings rest on a single LLM (GPT-4o-mini) with fixed simulation parameters including feed length, step count, and injection dosage. LLMs' position effects in long inputs \cite{liu2024lost} and their preferences for accurate information across personas \cite{liu-etal-2025-mosaic, chuang2024simulating} may inflate the upranking effect and overall engagement with verified news.

Finally, our human validation covers the injection treatments only, and the comparison study differs from our simulation in platform, measured behavior and duration, so we compare direction and dose-normalized efficiency rather than raw effect size. To our knowledge, no directly comparable human-subject study exists for upranking or celebrity source cues. Our validation therefore functions as initial evidence for subsequent human testing rather than as a matched human--agent replication. 

\section*{Ethical Considerations}
This study uses LLM agents in a sandbox simulation to compare candidate interventions at an early stage: no human participants are involved, no live recommender system is modified, the content pool consists of publicly posted tweets, and agent personas are synthetic. Thus, there are no major ethical considerations associated with our work as such. 

\section*{Acknowledgments}
The authors gratefully acknowledge the support of the National Science Foundation (Award 2427809, PI Magdalena Wojcieszak), the European Research Council (ERC Consolidator 101126218, NEWSUSE, PI Magdalena Wojcieszak) and of the Center for Excellence in Social Sciences at the University of Warsaw, which funded the project. Any opinions, findings, conclusions, or recommendations expressed in this material are those of the authors and do not necessarily reflect the views of the funders.

\bibliography{custom}

\clearpage

\appendix

\section{Dataset}
\label{app:dataset}
We construct a realistic simulation environment by integrating two data sources: a large-scale collection of general real-world Twitter/X posts and a collection of posts from high-quality news organizations. The final merged dataset contains 467,312 posts generated by 12,103 unique users. Each post in the dataset contains the original post content, author identities, engagement counts, and a synthetic timestamp used in the simulation feed.

\subsection{Data Source}
\label{app:data_source}
\paragraph{General Twitter/X Data.}
We collected 477,480 English-language real-world Twitter/X data using Apify. This source provides the general posts pool for approximating the background distribution of posts that users may encounter on Twitter/X. After preprocessing, this source contributes 435,273 original, text-only posts authored by 12,068 unique users.

\paragraph{High-Quality News Data.}
To control the quality of news intervention, we select 35 high-credibility news outlets (e.g., Reuters, AP, NPR. See Appendix \ref{app:news_list} for the full list). We use a publicly available Twitter/X dataset released on HuggingFace (Twitter100M) as the source of High-Quality News data. The dataset contains around 150K users and 90M tweets that collected in 2023. For each user, up to 1000 most recent tweets are recorded. In our study, we filter for tweets authored by the 35 high-quality news outlets. This yields a subset of 32,039 news posts.

\subsection{Preprocessing}
\label{app:preprocessing}
\paragraph{Data Filtering.} We exclude reposts, retweets, reshares, and comments from the dataset, as these posts depend heavily on prior conversational context. We retain only original tweets authored by users. We also exclude tweets containing external links or images, because agent inputs are limited to textual content.
\paragraph{Synthetic Timestamps.}
The raw timestamps of the source datasets span over years. To mimic an social media environment that contains most recent information, we reassign each post a synthetic timestamp within a compressed three-day window from 2023-05-01 to 2023-05-03. This reassignment is independent of the original creation time of the post. The goal is to make the posts shown in a generated feed appear temporally ``fresh'', as they would in a real browsing session.

\subsection{News and Political Classification}
\label{app:pol_class}
\paragraph{Definition of News.} In our main study, we label all posts from the 35 high-credibility news outlets as ``news'' across all conditions. 

\paragraph{Motivation of Political Content.} To simulate the realistic political ecology on Twitter/X, we control the proportion of political posts in a feed to 10\% \cite{pew2022}. We employ a classifier to label the political leaning of all non-news posts, while automatically identify all posts from high-quality news data as ``news''. This allows us to maintain a target ratio of political content during the generation of random recommendation sequence.

\paragraph{Definition of ``Political''}
\label{app:political_def}
We follow the broad conceptualization of political content in \citet{askari2024incentivizing} to define a post as political if it references political figures, policies, elections, or specific political events, as well as politically relevant social issues. We use the classifier reported in this paper as well to identify political posts. The classifier achieved 91.47\% accuracy, 90.34\% precision, and 90.77\% recall on a combined Facebook and Reddit test set of 3,885 manually annotated comments, and achieved 93.77\% accuracy, 92.25\% precision, and 92.95\% recall on Reddit.

\section{Prompts}

\subsection{Prompt for Twitter Agent Simulation}
\label{sec:twitter_agent_prompt}
The agents receive the following system and user prompts.
\begin{promptlisting}
(*@\textbf{System prompt}@*)
# OBJECTIVE
You are a Twitter user, and I will present you with some posts. After you see the posts, choose actions from the available functions. You may see posts from famous celebrities marked with (Verified).

# SELF-DESCRIPTION
Your actions should be consistent with your self-description and personality.
Your name is <NAME>.
Your username is <USERNAME>.
You are a <AGE>-year-old <GENDER>. Your education background is <EDUCATION>.

# RESPONSE METHOD
Please perform actions by tool calling.

(*@\textbf{User prompt}@*)
Here is your social media feed: after refreshing, you see some posts <POST_LIST>.
Read through those posts and engage with those you find relevant or interesting to your profile and based on the posts' content. Behave naturally, as you would on a real Twitter platform, and remember that you can like posts, comment on them, repost them, or follow the posting accounts.
\end{promptlisting}

\section{Agent Generation}
\label{app:agents}
\subsection{Demographic Distribution and Population Generation}
\label{app:demo}
To ensure the ecological validity of our simulation, we construct the agent population to mirror the demographic distribution of active U.S. Twitter users in 2024 \cite{Xeducation2024, Xgender2025, Xage2025}. Table \ref{tab:pop} shows the demographic distribution details. We employed a stratified probabilistic sampling approach. 
\begin{table}[ht]
\centering
\footnotesize
\setlength{\tabcolsep}{3pt}
\caption{Demographic distributions of the simulated population.}
\label{tab:pop}
\begin{tabular}{l >{\raggedright\arraybackslash}p{3.3cm} r}
\hline
\textbf{Dimension} & \textbf{Category} & \textbf{Probability (\%)} \\
\hline
\multirow{2}{*}{Gender}
    & Male   & 63.7 \\
    & Female & 36.3 \\
\hline
\multirow{5}{*}{Age}
    & 13--17 & 2.0 \\
    & 18--24 & 32.1 \\
    & 25--34 & 37.5 \\
    & 35--49 & 21.1 \\
    & 50--70 & 7.3 \\
\hline
\multirow{6}{*}{Education}
    & No formal education               & 10 \\
    & Some high school                  & 20 \\
    & Some college                      & 20 \\
    & Bachelor's degree or equivalent   & 20 \\
    & Master's degree or equivalent     & 15 \\
    & Doctoral degree or equivalent     & 15 \\
\hline
\end{tabular}
\end{table}

For education level above bachelor's or equivalent, common university majors will be provided randomly from the following 15 majors:
Business Administration, Psychology, Computer Science, Biology, Engineering, Communications, English, Economics, Political Science, Marketing, Education, Sociology, History, Mathematics, Finance.

To construct a large-scale population while ensuring statistical robustness, agents were generated in independent batches. For each instance, demographic attributes were sampled I.I.D. from the target distribution $\mathcal{D}_{pop}$. To ensure internal validity, agents were randomly assigned to experimental conditions (Control, Inject-2, Inject-4, Upranking, Celebrity) independent of their generation order. Given the I.I.D. nature of the generation process, this guarantees that all treatment groups are statistically homogeneous in their demographic distributions prior to any intervention, minimizing selection bias.

\subsection{Agent Profile Generation}
\label{app:profile}
\paragraph{Generation Logic.}
To ensure that the synthetic agents exhibit realistic Twitter-like digital self-presentation, we employed LLMs to create user profiles (Display Name and Handles) based on their different demographic attributes. For each agent, based on their stratified demographic attributes (Age, Gender, Education), we prompted the GPT-4o-mini (Temperature=0.8) to generate a Realname (Display Name) and a Username (Handle). 

\paragraph{Prompt Design.}
The prompt below instructs the model to vary naming patterns and avoid synthetic artifacts.
\begin{promptlisting}
You are an expert in Digital Sociology and Internet Linguistics. Your task is to synthesize realistic social media identities for a diverse population.

### Simulation Goal
The goal is to mimic the diverse naming patterns observed on real social media platforms. For each user description provided, infer their likely "Digital Fingerprint" based on their demographics (Age, Education, Background). Avoid generic or bot-like patterns.

### Generation Guidelines:
1. **REALNAME (Display Name):**
- Objective: Reflect the user's self-perception.
- Instruction: Ensure a mix of formatting styles. Some users use full legal names (e.g., "Jonathan Smith"), some use casual nicknames (e.g., "Jonny"), some include credentials (e.g., "Jon Smith, PhD"), and younger users often use stylistic lowercase or emojis (e.g., "jon (*@\leftmoon@*)").

2. **USERNAME (Handle):**
- Objective: Create a unique human-like handle, reflect the **era** in which they likely joined the internet.
- Style: Keep it between 8-15 characters and realistic for Twitter.
- Digital Literacy:
    - High literacy (often younger) = Abstract, aesthetic, short, or witty handles.
    - Low literacy (often older) = Formulaic, name+sequence, or rigid handles.
- Ban any numbers that can be inferred from the profile (e.g., age).
- Other numbers are allowed if they look natural for Twitter handles.

### Output Format:
CRITICAL: Return ONLY a valid JSON array.
[
  {"realname": "...", "username": "..."},
  ...
]
\end{promptlisting}

\paragraph{Profile Examples.}
Table \ref{tab:agent_samples} exhibits the diversity of profiles generated across different demographic groups.

\begin{table*}[h]
\centering
\small
\caption{Sample of generated agent identities demonstrating demographic coverage.}
\begin{tabular}{clclll}
\hline
\textbf{Age} & \textbf{Sex} & \textbf{Education} & \textbf{Display Name} & \textbf{Handle} \\ \hline
19 & M & Bachelor's & Liam O'Connor & @LiamPoliSci \\
28 & F & Master's & Emma R. & @psychology\_28 \\
38 & F & Bachelor's & Sarah Coder & @codewithsarah \\
40 & M & Ph.D. & Dr. David Lee & @psych\_david \\
55 & M & Bachelor's & Jonathan Smith & @JonnyWrites \\
15 & M & None & Tyler \smiley & @tyler\_no\_school \\
25 & M & Some high school & Jake the Great & @jake4real \\
\hline
\end{tabular}
\label{tab:agent_samples}
\end{table*}

\section{List of News Sources}
\label{app:news_list}
We compiled a list of high-quality news outlets to serve as the news source for our simulation. 
\paragraph{Aggregated news quality ratings.} We first obtain a candidate list of news domains from the public news-domain quality dataset released in \citet{lin2023high}. This resource aggregates domain quality assessments from six independent rating organizations and provides an aggregated quality score for each domain performed imputation together with principal component analysis. We then retain outlets whose aggregated quality score exceeds a threshold of 0.8. We conducted a manual sanity check of candidate outlets around the selected thresholds to ensure that the resulting outlet set corresponds to widely recognized professional news organizations, and exclude individual journalist or editor accounts and local news outlets.
\paragraph{Credibility and partisan bias from Ad Fontes Media.} We further filter candidate outlets using credibility and partisan bias assessments from Ad Fontes Media, an independent media research organization that provides human-coded evaluations of news outlets along two dimensions \cite{adfontes2025}. Specifically, we retain only outlets that with high credibility scores ($> 35$) and low bias scores ($-12$ to $12$).

Table \ref{tab:news} shows the full list of 35 selected news outlets, which are selected based on both aggregated news quality ratings and Ad Fontes credibility and partisan bias scores. Each entry includes the outlet's official Twitter handle (username), display name (name) and the reliability score.

\begin{table*}[t]
\centering
\caption{High-Quality News Outlets on Twitter.}
\small
\begin{tabular}{llrrr}
\hline
username & name & score & credibility & bias \\
\hline
CBSNews & CBS News & 0.880 & 41.99 & -2.86 \\
ForeignAffairs & Foreign Affairs & 0.909 & 43.58 & -7.11 \\
NPR & NPR & 0.931 & 43.08 & -4.31 \\
NewsNation & NewsNation & 0.947 & 42.95 & 0.53 \\
Reuters & Reuters & 1.000 & 44.98 & -1.40 \\
AP\_Politics & AP Politics & 0.998 & 44.80 & -2.51 \\
goodnewsnetwork & Good News Network & 0.932 & 42.57 & -1.43 \\
nprpolitics & NPR Politics & 0.931 & 43.08 & -4.31 \\
ForeignPolicy & Foreign Policy & 0.919 & 42.63 & -4.05 \\
USATODAY & USA TODAY & 0.901 & 40.81 & -3.73 \\
thehill & The Hill & 0.899 & 41.64 & -1.49 \\
axios & Axios & 0.897 & 43.05 & -3.31 \\
FortuneMagazine & FORTUNE & 0.876 & 36.96 & -2.91 \\
ABC & ABC News & 0.863 & 44.30 & -2.79 \\
ABCPolitics & ABC News Politics & 0.863 & 44.30 & -2.79 \\
ABCWorldNews & World News Tonight & 0.863 & 44.30 & -2.79 \\
politico & POLITICO & 0.863 & 42.33 & -5.41 \\
nytimes & The New York Times & 0.859 & 40.93 & -8.04 \\
propublica & ProPublica & 0.853 & 47.22 & -6.82 \\
CNBC & CNBC & 0.850 & 43.75 & -1.95 \\
CNBCnow & CNBC Now & 0.850 & 43.75 & -1.95 \\
thisisinsider & Insider & 0.847 & 40.66 & -3.96 \\
NBCPolitics & NBC Politics & 0.844 & 42.86 & -5.37 \\
NBCNews & NBC News & 0.844 & 42.86 & -5.37 \\
vulture & Vulture & 0.843 & 39.42 & -7.49 \\
bpolitics & Bloomberg Politics & 0.840 & 42.29 & -2.93 \\
BostonGlobe & The Boston Globe & 0.835 & 41.94 & -7.85 \\
Forbes & Forbes & 0.830 & 40.74 & -2.86 \\
TIME & TIME & 0.826 & 41.10 & -7.37 \\
washingtonpost & The Washington Post & 0.819 & 38.67 & -7.01 \\
Newsweek & Newsweek & 0.818 & 38.73 & -1.80 \\
CBC & CBC & 0.817 & 44.39 & -5.01 \\
BusinessInsider & Business Insider & 0.805 & 40.66 & -3.96 \\
WSJ & The Wall Street Journal & 0.804 & 43.09 & 4.39 \\
TheAtlantic & The Atlantic & 0.804 & 37.80 & -9.37 \\
\hline
\label{tab:news}
\end{tabular}
\end{table*}

\section{List of Celebrities}
\label{app:celebrity}
To construct the celebrity pool for the Celebrity treatment, we aimed to select celebrities with maximal potential network impact. We sourced the top 50 most-followed Twitter accounts in 2025 \cite{Top50followedTwitter}, and identified all individuals (rather than organizations or companies) who primarily post in English. From this list, we further removed all politicians, government officials, and other political figures (e.g., Elon Musk) to isolate cultural influence from business or political authority. From the remaining candidates, we selected the top 10 accounts, balanced equally by gender (5 male and 5 female). Table \ref{tab:celeb} represents the full list of celebrities used in our simulation. 

\begin{table}[ht]
\centering
\caption{Selected Most Followed Celebrities on Twitter.}
\begin{tabular}{lll}
\hline
realname & username & gender \\
\hline
Justin Bieber & justinbieber & male \\
Justin Timberlake & jtimberlake & male \\
LeBron James & KingJames & male \\
Bruno Mars & BrunoMars & male \\
Niall Horan & NiallOfficial & male \\
Rihanna & rihanna & female \\
Katy Perry & katyperry & female \\
Taylor Swift & taylorswift13 & female \\
Lady Gaga & ladygaga & female \\
Kim Kardashian & KimKardashian & female \\
\hline
\end{tabular}
\label{tab:celeb}
\end{table}

\section{Integration with the OASIS Framework}
\label{app:oasis}
We implemented our simulation pipeline on top of the OASIS framework \cite{yang2024oasis}. OASIS provides the underlying infrastructure for large-scale agent execution, multi-round simulation, LLM-based action through function calling, and persistent agent memory across time steps. In our study, we extend and modify the OASIS framework to support the specific requirements of our experimental setting, including custom agent construction, post pool construction, feed generation, and intervention mechanisms.

While the original OASIS framework enables the support of networked interactions and emergent network effects, our study does not emphasize network dynamics. Instead, we focus on how agents react to content exposure under different interventions. Accordingly, we restrict the agent action space to \textit{like}, \textit{comment}, \textit{repost}, and \textit{follow}, and do not allow agents to create new posts. In turn, all agents are exposed to feeds sampled from a fixed post pool initialized prior to the simulation for post pool construction and feed generation (see Appendix~\ref{app:feed} for details of post dataset and feed construction method).
Our system implements all task-specific components and experimental logic on top of OASIS to support this experimental setting. At each simulation time step, our OASIS related pipeline proceeds as follows:
\begin{enumerate}
    \item construct a baseline recommendation feed from the post pool;
    \item apply specific feed transformations based on the treatment type;
    \item invoke agent actions through LLM function calls and update agent memory;
    \item record agent actions and system states for evaluation.
\end{enumerate}
The detailed implementation of each component is described in the subsequent sections of this Appendix.

\section{Simulation Timeline and Memory Management}
\label{app:timeline}
Our simulation runs for $T=20$ discrete time steps. To mimic the limited working memory of human cognition, we implemented a \textit{Sliding Window Memory} mechanism. At the end of each step, the framework truncates the agent's memory to retain only the most recent 5 interaction steps. This ensures that the agent's decision-making is driven by immediate context, preventing context saturation while maintaining temporal memory. The memory contains the feeds and action history generated in previous steps, allowing the agent to maintain consistency in its persona.

\section{RecSys Logic: Baseline Random Feed Generation}
\label{app:feed}
To construct a realistic information environment, we explicitly control the proportion of political content in user feed. A Pew Research Center analysis indicates that approximately 8\% of English-language original tweets posted between May 1, 2020, and May 31, 2021  are political \cite{pew2022}. Thus, we use 10\% as the expected average proportion of political content in the feeds to ensure sufficient signal for our study. 

This sampling is implemented by a Binomial Sampling Scheme. For a feed of length $N=40$, the number of political posts $K$ is drawn from a binomial distribution:
$$K \sim \text{Binomial}(n=N, p=0.1)$$
This ensures that the expected political proportion is $E[K]/N=0.1$, while allowing natural variation across different users and time steps.

Given $K$, we sampled $K$ political posts and $N-K$ non-political posts uniformly without replacement from their respective content pools. Sampled posts were then sorted by their created timestamps (from latest to oldest) before being shown to the agent.

\section{Implementation Details of Treatments}
\label{app:treatment}
\subsection{News Injection}
Algorithm \ref{alg:injection} describes the news injection procedure parameterized by injection count $k_{inject} \in \{2, 4\}$. At each time step, the system first randomly samples 4 news posts from the news pool as a shared pre-selected payload. Inject-2 uses the first two posts, Inject-4 uses all four. The algorithm then randomly identifies $k_{inject}$ posts in the current recommendation sequence and replaces them with the selected news posts. This method preserves the total number of posts in the feed presented to the agent.

\begin{algorithm}[ht]
\caption{Content Injection Treatment ($\mathcal{T}_{inject}$)}
\label{alg:injection}

\textbf{Input}: Recommendation sequence $\mathcal{R}$; news set $\mathcal{N}$ \\
\textbf{Parameter}: Injection count $k \in \{2, 4\}$\\
\textbf{Output}: User feed $\mathcal{F}$

\begin{algorithmic}[1]
    \STATE $S_{pre} = \text{Sample}(\mathcal{N}, 4)$ \COMMENT{Shared payload, fixed per time step}
    \STATE $S_{news} = S_{pre}[1:k_{inject}]$ \COMMENT{First 2 if $k_{inject}{=}2$; all 4 if $k_{inject}{=}4$}
    \STATE Let $\mathcal{F} = \text{Copy}(\mathcal{R})$
    \STATE $I_{cand} = \{ i \mid F[i] \notin S_{news} \}$
    \STATE $I_{replace} = \text{Sample}(I_{cand}, k_{inject})$
    \FOR{$j = 0$ \TO $k_{inject}-1$}
        \STATE $F[I_{replace}[j]] = S_{news}[j]$
    \ENDFOR
    \STATE \textbf{return} $\mathcal{F}$
\end{algorithmic}
\end{algorithm}

\subsection{Upranking}
Algorithm \ref{alg:upranking} takes the injected feed produced by Algorithm \ref{alg:injection} (with $k_{inject}{=}2$) and promotes the two injected news posts to target positions 0 (1st post) and index 2 (3rd post) to mimic a ``top-of-feed'' positional bias while allowing for one non-injected post in between (i.e., 2nd post). For each injected news post not already at its target position, the algorithm swaps it with the post currently occupying that slot.

\begin{algorithm}[ht]
\caption{Positional Upranking Treatment ($\mathcal{T}_{rank}$)}
\label{alg:upranking}
\textbf{Input}: Injected feed $\mathcal{F}$ (output of Algorithm \ref{alg:injection} with $k_{inject}{=}2$); injected news set $S_{news}$\\
\textbf{Parameter}: Target positions $P_{top} = [0, 2]$\\
\textbf{Output}: Upranked feed $\mathcal{F}$

\begin{algorithmic}[1]
    \STATE $slot = 0$
    \FOR{each news post $n$ in $S_{news}$, in order of position in $\mathcal{F}$}
        \IF{$slot \geq |P_{top}|$}
            \STATE \textbf{break}
        \ENDIF
        \STATE $idx_n = \text{IndexOf}(\mathcal{F}, n)$
        \IF{$idx_n \neq P_{top}[slot]$}
            \STATE $\text{Swap}(\mathcal{F}[P_{top}[slot]],\; \mathcal{F}[idx_n])$ \COMMENT{Move news to target position}
        \ENDIF
        \STATE $slot = slot + 1$
    \ENDFOR
    \STATE \textbf{return} $\mathcal{F}$
\end{algorithmic}
\end{algorithm}

\subsection{Celebrity Source Rendering}
Algorithm \ref{alg:celebrity} takes the injected feed produced by Algorithm \ref{alg:injection} (with $k_{inject}{=}2$). It shows the logic for source cue intervention. Unlike Upranking and Injection, which alter the quantity or structural arrangement of the feed, this treatment operates only on the \textit{presentation layers}. The procedure iterates through the feed. For each injected news post, the algorithm samples a celebrity from the celebrity pool $\mathcal{C}$. Then, overwrits the visible author attributes with those of the sampled celebrity. All other attributes, including content, position, and non-news posts, remain unchanged. 

\begin{algorithm}[ht]
\caption{Celebrity Source Masking ($\mathcal{T}_{source}$)}
\label{alg:celebrity}
\textbf{Input}: Injected feed $\mathcal{F}$ (output of Algorithm \ref{alg:injection} with $k_{inject}{=}2$); injected news set $S_{news}$; celebrity pool $\mathcal{C}$\\
\textbf{Output}: Display-modified feed $\mathcal{F}$

\begin{algorithmic}[1]
    \FOR{each post $p$ in $\mathcal{F}$}
        \IF{$p \in S_{news}$}
            \STATE $c = \text{Sample}(\mathcal{C})$ \COMMENT{One celebrity per news post}
            \STATE $p.\text{name} \leftarrow c.\text{realname} + \text{``(Verified)''}$
            \STATE $p.\text{handle} \leftarrow c.\text{username}$
        \ENDIF
    \ENDFOR
    \STATE \textbf{return} $\mathcal{F}$
\end{algorithmic}
\end{algorithm}

\section{Code and Reproducibility}
\label{app:code}
To ensure the replicability of our work, we open-sourced the complete simulation pipeline, including the agent environment setup, prompt templates, and statistical analysis scripts. These materials are accessible at: \url{https://anonymous.4open.science/r/intervention-7723/}.

\
\section{Additional Results}
\subsection{Distribution of News and Political Posts}
\label{app:distribution_news_political}
Table \ref{tab:perslate_feed_distribution} reports the distribution of news and political posts in the simulated 40-post feeds by treatment conditions. The definition of ``news'' and ``political'' are described in Appendix \ref{app:pol_class}. The reported values are computed at the per-user-per-step level, consistent with the unit used in the GLMM analyses. Although the news posts are sparse in our dataset, the control condition can still contain some news because feed sampling preserves the political composition of the post pool, and all news posts are also classified as political posts. The table therefore provides a descriptive check of feed composition across treatments before modeling engagement outcomes.

\begin{table*}[t]
  \centering
  \caption{Per-user-per-step distribution of news and political posts in the 40-post feed, by treatment. Each user-step is one observation. 13,835 agents $\times $20 simulation steps leads to 276,700 per-user-per-step observations. The engagement models use 276,589; we exclude the 111 agent-steps (0.04\%) because they contain no valid actions for analysis and their omission does not affect the estimates. Mean and SD with cluster-bootstrapped 95\% CI (500 resamples), Median and the [P5, P95] percentiles describe the empirical distribution.}
  \label{tab:perslate_feed_distribution}
  \small
  \setlength{\tabcolsep}{4pt}
  \begin{tabular}{lcccc}
  \toprule
  \textbf{Treatment} & \textbf{Mean $\pm$ SD} & \textbf{95\% CI (cluster boot)} & \textbf{Median} & \textbf{[P5, P95]} \\
  \midrule
  \multicolumn{5}{l}{\textit{Panel A --- News posts per 40-post feed}} \\[2pt]
  Control & $0.517 \pm 0.713$ & $[0.511,\ 0.522]$ & 0 & $[0,\ 2]$ \\
  Inject-2 & $2.489 \pm 0.696$ & $[2.484,\ 2.495]$ & 2 & $[2,\ 4]$ \\
  Inject-4 & $4.462 \pm 0.677$ & $[4.456,\ 4.467]$ & 4 & $[4,\ 6]$ \\
  Celebrity & $2.487 \pm 0.695$ & $[2.481,\ 2.494]$ & 2 & $[2,\ 4]$ \\
  Upranking & $2.489 \pm 0.692$ & $[2.483,\ 2.494]$ & 2 & $[2,\ 4]$ \\
  \addlinespace[6pt]
  \multicolumn{5}{l}{\textit{Panel B --- Political posts per 40-post feed}} \\[2pt]
  Control & $3.997 \pm 1.894$ & $[3.980,\ 4.014]$ & 4 & $[2,\ 7]$ \\
  Inject-2 & $5.775 \pm 1.846$ & $[5.761,\ 5.790]$ & 6 & $[4,\ 9]$ \\
  Inject-4 & $7.602 \pm 1.809$ & $[7.586,\ 7.618]$ & 7 & $[5,\ 11]$ \\
  Celebrity & $5.787 \pm 1.851$ & $[5.773,\ 5.801]$ & 6 & $[4,\ 9]$ \\
  Upranking & $5.793 \pm 1.851$ & $[5.776,\ 5.808]$ & 6 & $[4,\ 9]$ \\
  \bottomrule
  \end{tabular}
\end{table*}

\subsection{Alternative Measure: News Engagement Rate}
\label{app:rate_model}
Table \ref{tab:rate_poisson}, \ref{tab:lmm_rate} and \ref{tab:nb2_rate} display the full statistical results of Poisson Mixed-Effects, Linear Mixed-Effects and Negative Binomial with news engagement rate outcomes.

\begin{table*}[!ht]
\centering
\caption{Poisson GLMM with Offset (Rate Model) of News Engagement. The dependent variable is the news engagement count with $\log(\text{total\_activity})$ as offset, yielding a model for the share of activity that is news-directed. Fit on user-steps with $\text{total\_activity} > 0$. Coefficients on the log scale; standard errors in parentheses. Intervention effects are relative to Control; demographic effects are relative to the reference categories (Female for gender; Bachelor's degree for education). All models include a user-level random intercept. *** $p < 0.001$, ** $p < 0.01$, * $p < 0.05$.}
\label{tab:rate_poisson}
\scriptsize
\setlength{\tabcolsep}{3pt}
\begin{tabular}{lccccc}
\toprule
 & \textbf{Model 1} & \textbf{Model 2} & \textbf{Model 3} & \textbf{Model 4} & \textbf{Model 5} \\
\textbf{Variable} & \textbf{Total} & \textbf{Likes} & \textbf{Comments} & \textbf{Reposts} & \textbf{Follows} \\
\midrule
Intercept & $-4.487^{***}$ (0.028) & $-5.138^{***}$ (0.038) & $-6.555^{***}$ (0.077) & $-6.228^{***}$ (0.067) & $-7.599^{***}$ (0.102) \\
\addlinespace[0.5em]
\multicolumn{6}{l}{\textit{Treatments (Ref: Control)}} \\
Celebrity          & $2.641^{***}$ (0.028) & $2.096^{***}$ (0.039) & $2.986^{***}$ (0.077) & $3.199^{***}$ (0.066) & $3.092^{***}$ (0.097) \\
Inject-2           & $1.964^{***}$ (0.029) & $1.827^{***}$ (0.039) & $2.647^{***}$ (0.077) & $1.970^{***}$ (0.068) & $1.234^{***}$ (0.103) \\
Inject-4           & $2.423^{***}$ (0.028) & $2.313^{***}$ (0.038) & $2.993^{***}$ (0.076) & $2.505^{***}$ (0.067) & $1.765^{***}$ (0.100) \\
Upranking          & $3.643^{***}$ (0.028) & $3.834^{***}$ (0.037) & $3.722^{***}$ (0.075) & $3.003^{***}$ (0.066) & $2.964^{***}$ (0.096) \\
\addlinespace[0.5em]
\multicolumn{6}{l}{\textit{Controls (Ref: Female; Bachelor's degree)}} \\
Age (Centered)     & $0.002^{***}$ (0.000) & $0.001^{***}$ (0.000) & $0.005^{***}$ (0.001) & $0.003^{**}$ (0.001) & $-0.012^{***}$ (0.002) \\
Gender: Male       & $-0.043^{***}$ (0.007) & $-0.018^{*}$ (0.009) & $-0.079^{***}$ (0.020) & $-0.091^{***}$ (0.021) & $0.005$ (0.040) \\
Edu: Doctoral      & $0.056^{***}$ (0.013) & $0.041^{**}$ (0.016) & $0.137^{***}$ (0.036) & $0.033$ (0.038) & $-0.201^{*}$ (0.078) \\
Edu: Master's      & $0.013$ (0.011) & $0.008$ (0.014) & $0.065^{*}$ (0.032) & $-0.059$ (0.033) & $0.034$ (0.065) \\
Edu: Some college  & $-0.074^{***}$ (0.010) & $-0.049^{***}$ (0.013) & $-0.188^{***}$ (0.030) & $-0.130^{***}$ (0.031) & $0.181^{**}$ (0.058) \\
Edu: Some HS       & $-0.085^{***}$ (0.010) & $-0.020$ (0.012) & $-0.269^{***}$ (0.030) & $-0.211^{***}$ (0.031) & $0.125^{*}$ (0.058) \\
Edu: No formal     & $-0.097^{***}$ (0.012) & $0.002$ (0.015) & $-0.334^{***}$ (0.037) & $-0.271^{***}$ (0.037) & $0.066$ (0.069) \\
\midrule
Observations       & 273{,}134 & 273{,}134 & 273{,}134 & 273{,}134 & 273{,}134 \\
Users              & 13{,}835  & 13{,}835  & 13{,}835  & 13{,}835  & 13{,}835 \\
\bottomrule
\end{tabular}
\end{table*}

\subsection{Robustness Check: Alternative Models}
\label{app:full_stats}
\paragraph{Full Results of Linear Mixed-Effects Models.}
\begin{equation}
  \begin{split}
  Y_{it} =
  &\ \beta_0 + \beta_1 \text{Treat}_i + \beta_2 \text{Age}_i \\
  &+ \beta_3 \text{Gender}_i + \beta_4 \text{Edu}_i + u_i + \varepsilon_{it},
  \end{split}
  \label{eq:lmm_count}
  \end{equation}
We do not rely on the LMM as our primary model because the engagement outcomes are non-negative ``count'' values with heavy right skew, violating the normality assumption of Gaussian residuals. Table \ref{tab:lmm_count} and \ref{tab:lmm_rate} shows the full results of news engagement count and news engagement rate separately.
% ======== Updated Tables ==========
\begin{table*}[!ht]
\centering
\caption{Linear Mixed Model (Gaussian) of News Engagement Count. The dependent variable is the news engagement count per user per step. Coefficients are on the response scale (counts per step); standard errors in parentheses. Intervention effects are relative to Control; demographic effects are relative to the reference categories (Female for gender; Bachelor's degree for education). All models include a user-level random intercept. *** $p < 0.001$, ** $p < 0.01$, * $p < 0.05$.}
\label{tab:lmm_count}
\scriptsize
\setlength{\tabcolsep}{3pt}
\begin{tabular}{lccccc}
\toprule
 & \textbf{Model 1} & \textbf{Model 2} & \textbf{Model 3} & \textbf{Model 4} & \textbf{Model 5} \\
\textbf{Variable} & \textbf{Total} & \textbf{Likes} & \textbf{Comments} & \textbf{Reposts} & \textbf{Follows} \\
\midrule
Intercept & $0.060^{***}$ (0.006) & $0.020^{***}$ (0.005) & $0.020^{***}$ (0.003) & $0.019^{***}$ (0.003) & $0.001$ (0.002) \\
\addlinespace[0.5em]
\multicolumn{6}{l}{\textit{Treatments (Ref: Control)}} \\
Celebrity          & $0.334^{***}$ (0.006) & $0.101^{***}$ (0.005) & $0.070^{***}$ (0.003) & $0.113^{***}$ (0.003) & $0.050^{***}$ (0.002) \\
Inject-2           & $0.214^{***}$ (0.006) & $0.104^{***}$ (0.005) & $0.060^{***}$ (0.003) & $0.040^{***}$ (0.003) & $0.010^{***}$ (0.002) \\
Inject-4           & $0.360^{***}$ (0.006) & $0.182^{***}$ (0.005) & $0.086^{***}$ (0.003) & $0.073^{***}$ (0.003) & $0.019^{***}$ (0.002) \\
Upranking          & $1.601^{***}$ (0.006) & $1.065^{***}$ (0.005) & $0.251^{***}$ (0.003) & $0.176^{***}$ (0.003) & $0.109^{***}$ (0.002) \\
\addlinespace[0.5em]
\multicolumn{6}{l}{\textit{Controls (Ref: Female; Bachelor's degree)}} \\
Age (Centered)     & $0.001^{***}$ (0.000) & $0.000^{**}$ (0.000) & $0.001^{***}$ (0.000) & $0.000^{***}$ (0.000) & $-0.000^{***}$ (0.000) \\
Gender: Male       & $-0.021^{***}$ (0.004) & $-0.005$ (0.003) & $-0.009^{***}$ (0.002) & $-0.007^{***}$ (0.002) & $0.000$ (0.001) \\
Edu: Doctoral      & $0.033^{***}$ (0.008) & $0.016^{*}$ (0.006) & $0.018^{***}$ (0.004) & $0.005$ (0.004) & $-0.006^{*}$ (0.003) \\
Edu: Master's      & $0.023^{***}$ (0.006) & $0.014^{*}$ (0.005) & $0.010^{**}$ (0.003) & $-0.003$ (0.003) & $0.001$ (0.002) \\
Edu: Some college  & $-0.047^{***}$ (0.006) & $-0.017^{***}$ (0.005) & $-0.021^{***}$ (0.003) & $-0.013^{***}$ (0.003) & $0.005^{*}$ (0.002) \\
Edu: Some HS       & $-0.051^{***}$ (0.006) & $-0.010^{*}$ (0.005) & $-0.027^{***}$ (0.003) & $-0.017^{***}$ (0.003) & $0.003$ (0.002) \\
Edu: No formal     & $-0.042^{***}$ (0.007) & $0.005$ (0.006) & $-0.029^{***}$ (0.003) & $-0.021^{***}$ (0.003) & $0.003$ (0.003) \\
\midrule
Observations       & 276{,}589 & 276{,}589 & 276{,}589 & 276{,}589 & 276{,}589 \\
Users              & 13{,}835  & 13{,}835  & 13{,}835  & 13{,}835  & 13{,}835 \\
\bottomrule
\end{tabular}
\end{table*}
\begin{table*}[!h]
\centering
\caption{Linear Mixed Model (Gaussian) on News Engagement Rate. The dependent variable is the per-step ratio of news-directed actions to total actions ($\text{news\_count}/\text{total\_activity}$). Fit on user-steps with $\text{total\_activity} > 0$. Coefficients are on the response (proportion) scale; standard errors in parentheses. Intervention effects are relative to Control; demographic effects are relative to the reference categories (Female for gender; Bachelor's degree for education). All models include a user-level random intercept. *** $p < 0.001$, ** $p < 0.01$, * $p < 0.05$.}
\label{tab:lmm_rate}
\scriptsize
\setlength{\tabcolsep}{3pt}
\begin{tabular}{lccccc}
\toprule
 & \textbf{Model 1} & \textbf{Model 2} & \textbf{Model 3} & \textbf{Model 4} & \textbf{Model 5} \\
\textbf{Variable} & \textbf{Total} & \textbf{Likes} & \textbf{Comments} & \textbf{Reposts} & \textbf{Follows} \\
\midrule
Intercept & $0.014^{***}$ (0.001) & $0.006^{***}$ (0.001) & $0.004^{***}$ (0.001) & $0.004^{***}$ (0.001) & $0.000$ (0.001) \\
\addlinespace[0.5em]
\multicolumn{6}{l}{\textit{Treatments (Ref: Control)}} \\
Celebrity          & $0.123^{***}$ (0.001) & $0.049^{***}$ (0.001) & $0.022^{***}$ (0.001) & $0.038^{***}$ (0.001) & $0.015^{***}$ (0.001) \\
Inject-2           & $0.045^{***}$ (0.001) & $0.021^{***}$ (0.001) & $0.013^{***}$ (0.001) & $0.009^{***}$ (0.001) & $0.002^{***}$ (0.001) \\
Inject-4           & $0.076^{***}$ (0.001) & $0.038^{***}$ (0.001) & $0.018^{***}$ (0.001) & $0.016^{***}$ (0.001) & $0.004^{***}$ (0.001) \\
Upranking          & $0.427^{***}$ (0.001) & $0.325^{***}$ (0.001) & $0.046^{***}$ (0.001) & $0.034^{***}$ (0.001) & $0.022^{***}$ (0.001) \\
\addlinespace[0.5em]
\multicolumn{6}{l}{\textit{Controls (Ref: Female; Bachelor's degree)}} \\
Age (Centered)     & $0.000^{***}$ (0.000) & $0.000^{**}$ (0.000) & $0.000^{***}$ (0.000) & $0.000^{**}$ (0.000) & $-0.000^{***}$ (0.000) \\
Gender: Male       & $-0.006^{***}$ (0.001) & $-0.002^{**}$ (0.001) & $-0.002^{***}$ (0.000) & $-0.002^{***}$ (0.000) & $-0.000$ (0.000) \\
Edu: Doctoral      & $0.008^{***}$ (0.002) & $0.003^{*}$ (0.001) & $0.004^{***}$ (0.001) & $0.002^{*}$ (0.001) & $-0.001^{*}$ (0.001) \\
Edu: Master's      & $0.003^{*}$ (0.001) & $0.001$ (0.001) & $0.002^{**}$ (0.001) & $-0.000$ (0.001) & $0.000$ (0.001) \\
Edu: Some college  & $-0.008^{***}$ (0.001) & $-0.003^{**}$ (0.001) & $-0.004^{***}$ (0.001) & $-0.003^{***}$ (0.001) & $0.001^{*}$ (0.000) \\
Edu: Some HS       & $-0.009^{***}$ (0.001) & $-0.001$ (0.001) & $-0.005^{***}$ (0.001) & $-0.003^{***}$ (0.001) & $0.001$ (0.000) \\
Edu: No formal     & $-0.008^{***}$ (0.001) & $0.000$ (0.001) & $-0.005^{***}$ (0.001) & $-0.004^{***}$ (0.001) & $0.001$ (0.000) \\
\midrule
Observations       & 273{,}134 & 273{,}134 & 273{,}134 & 273{,}134 & 273{,}134 \\
Users              & 13{,}835  & 13{,}835  & 13{,}835  & 13{,}835  & 13{,}835 \\
\bottomrule
\end{tabular}
\end{table*}

\paragraph{Full Results of Negative Binomial.}
\begin{equation}
  \begin{split}
  \log(\mu_{it}) =
  &\ \beta_0 + \beta_1 \text{Treat}_i + \beta_2 \text{Age}_i \\
  &+ \beta_3 \text{Gender}_i + \beta_4 \text{Edu}_i + u_i,
  \end{split}
  \label{eq:nb2_count}
  \end{equation}
The substantive conclusions are unchanged when using Negative Binomial Model. However, the estimated NB2 dispersion parameters are extremely large in several models and in some engagement metric outcomes (Likes and Follows) the NB2 fit reports undefined information criteria, indicating limited practical gain and occasional numerical instability. Table \ref{tab:nb2_count} and \ref{tab:nb2_rate} report the count- and rate-basic results
\begin{table*}[!ht]
\centering
\caption{Negative Binomial (NB2) GLMM of News Engagement Count. Specification matches Table~\ref{tab:main_results} but uses an \texttt{nbinom2} family. The dispersion parameter approached infinity for every metric, indicating no overdispersion; coefficients converge to the Poisson limit. Coefficients on the log scale; standard errors in parentheses. *** $p < 0.001$, ** $p < 0.01$, * $p < 0.05$.}
\label{tab:nb2_count}
\scriptsize
\setlength{\tabcolsep}{3pt}
\begin{tabular}{lccccc}
\toprule
 & \textbf{Model 1} & \textbf{Model 2} & \textbf{Model 3} & \textbf{Model 4} & \textbf{Model 5} \\
\textbf{Variable} & \textbf{Total} & \textbf{Likes} & \textbf{Comments} & \textbf{Reposts} & \textbf{Follows} \\
\midrule
Intercept & $-3.562^{***}$ (0.030) & $-4.314^{***}$ (0.042) & $-5.556^{***}$ (0.075) & $-5.237^{***}$ (0.066) & $-6.550^{***}$ (0.099) \\
\addlinespace[0.5em]
\multicolumn{6}{l}{\textit{Treatments (Ref: Control)}} \\
Celebrity          & $2.560^{***}$ (0.030) & $2.022^{***}$ (0.043) & $2.893^{***}$ (0.075) & $3.085^{***}$ (0.065) & $2.962^{***}$ (0.094) \\
Inject-2           & $2.136^{***}$ (0.030) & $1.957^{***}$ (0.043) & $2.755^{***}$ (0.076) & $2.099^{***}$ (0.067) & $1.350^{***}$ (0.100) \\
Inject-4           & $2.605^{***}$ (0.029) & $2.453^{***}$ (0.042) & $3.115^{***}$ (0.075) & $2.637^{***}$ (0.066) & $1.886^{***}$ (0.097) \\
Upranking          & $4.100^{***}$ (0.029) & $4.395^{***}$ (0.041) & $4.130^{***}$ (0.074) & $3.425^{***}$ (0.065) & $3.365^{***}$ (0.093) \\
\addlinespace[0.5em]
\multicolumn{6}{l}{\textit{Controls (Ref: Female; Bachelor's degree)}} \\
Age (Centered)     & $0.002^{***}$ (0.000) & $0.002^{***}$ (0.001) & $0.005^{***}$ (0.001) & $0.003^{***}$ (0.001) & $-0.011^{***}$ (0.002) \\
Gender: Male       & $-0.054^{***}$ (0.009) & $-0.048^{**}$ (0.015) & $-0.081^{***}$ (0.019) & $-0.090^{***}$ (0.019) & $0.000$ (0.038) \\
Edu: Doctoral      & $0.076^{***}$ (0.017) & $0.078^{**}$ (0.028) & $0.143^{***}$ (0.034) & $0.031$ (0.035) & $-0.187^{*}$ (0.074) \\
Edu: Master's      & $0.051^{***}$ (0.014) & $0.049^{*}$ (0.024) & $0.092^{**}$ (0.029) & $-0.032$ (0.031) & $0.053$ (0.061) \\
Edu: Some college  & $-0.103^{***}$ (0.013) & $-0.100^{***}$ (0.022) & $-0.205^{***}$ (0.028) & $-0.149^{***}$ (0.028) & $0.156^{**}$ (0.055) \\
Edu: Some HS       & $-0.112^{***}$ (0.013) & $-0.063^{**}$ (0.022) & $-0.287^{***}$ (0.028) & $-0.219^{***}$ (0.029) & $0.102$ (0.055) \\
Edu: No formal     & $-0.089^{***}$ (0.016) & $0.005$ (0.026) & $-0.325^{***}$ (0.034) & $-0.263^{***}$ (0.035) & $0.064$ (0.066) \\
\midrule
Observations       & 276{,}589 & 276{,}589 & 276{,}589 & 276{,}589 & 276{,}589 \\
Users              & 13{,}835  & 13{,}835  & 13{,}835  & 13{,}835  & 13{,}835 \\
\bottomrule
\end{tabular}
\end{table*}

\begin{table*}[!h]
\centering
\caption{Negative Binomial (NB2) GLMM with Offset (Rate Model). Specification matches Table~\ref{tab:rate_poisson} but uses an \texttt{nbinom2} family. The dispersion parameter approached infinity for every metric, indicating no overdispersion; coefficients converge to the Poisson-with-offset limit. Coefficients on the log scale; standard errors in parentheses. *** $p < 0.001$, ** $p < 0.01$, * $p < 0.05$.}
\label{tab:nb2_rate}
\scriptsize
\setlength{\tabcolsep}{3pt}
\begin{tabular}{lccccc}
\toprule
 & \textbf{Model 1} & \textbf{Model 2} & \textbf{Model 3} & \textbf{Model 4} & \textbf{Model 5} \\
\textbf{Variable} & \textbf{Total} & \textbf{Likes} & \textbf{Comments} & \textbf{Reposts} & \textbf{Follows} \\
\midrule
Intercept & $-4.487^{***}$ (0.028) & $-5.138^{***}$ (0.038) & $-6.555^{***}$ (0.077) & $-6.228^{***}$ (0.067) & $-7.599^{***}$ (0.102) \\
\addlinespace[0.5em]
\multicolumn{6}{l}{\textit{Treatments (Ref: Control)}} \\
Celebrity          & $2.641^{***}$ (0.028) & $2.096^{***}$ (0.039) & $2.986^{***}$ (0.077) & $3.199^{***}$ (0.066) & $3.092^{***}$ (0.097) \\
Inject-2           & $1.964^{***}$ (0.029) & $1.827^{***}$ (0.039) & $2.647^{***}$ (0.077) & $1.970^{***}$ (0.068) & $1.234^{***}$ (0.103) \\
Inject-4           & $2.423^{***}$ (0.028) & $2.313^{***}$ (0.038) & $2.993^{***}$ (0.076) & $2.505^{***}$ (0.067) & $1.765^{***}$ (0.100) \\
Upranking          & $3.643^{***}$ (0.028) & $3.834^{***}$ (0.037) & $3.722^{***}$ (0.075) & $3.003^{***}$ (0.066) & $2.964^{***}$ (0.096) \\
\addlinespace[0.5em]
\multicolumn{6}{l}{\textit{Controls (Ref: Female; Bachelor's degree)}} \\
Age (Centered)     & $0.002^{***}$ (0.000) & $0.001^{***}$ (0.000) & $0.005^{***}$ (0.001) & $0.003^{**}$ (0.001) & $-0.012^{***}$ (0.002) \\
Gender: Male       & $-0.043^{***}$ (0.007) & $-0.018^{*}$ (0.009) & $-0.079^{***}$ (0.020) & $-0.091^{***}$ (0.021) & $0.005$ (0.040) \\
Edu: Doctoral      & $0.056^{***}$ (0.013) & $0.041^{*}$ (0.016) & $0.137^{***}$ (0.036) & $0.033$ (0.038) & $-0.201^{*}$ (0.078) \\
Edu: Master's      & $0.013$ (0.011) & $0.008$ (0.014) & $0.065^{*}$ (0.032) & $-0.059$ (0.033) & $0.034$ (0.065) \\
Edu: Some college  & $-0.074^{***}$ (0.010) & $-0.049^{***}$ (0.013) & $-0.188^{***}$ (0.030) & $-0.130^{***}$ (0.031) & $0.181^{**}$ (0.058) \\
Edu: Some HS       & $-0.085^{***}$ (0.010) & $-0.020$ (0.012) & $-0.269^{***}$ (0.030) & $-0.211^{***}$ (0.031) & $0.125^{*}$ (0.058) \\
Edu: No formal     & $-0.097^{***}$ (0.012) & $0.002$ (0.015) & $-0.334^{***}$ (0.037) & $-0.271^{***}$ (0.037) & $0.066$ (0.069) \\
\midrule
Observations       & 273{,}134 & 273{,}134 & 273{,}134 & 273{,}134 & 273{,}134 \\
Users              & 13{,}835  & 13{,}835  & 13{,}835  & 13{,}835  & 13{,}835 \\
\bottomrule
\end{tabular}
\end{table*}

\subsection{Robustness Check: Engagement with Political Posts}
\label{app:robust_political_engagement}
In our main paper, we define ``news posts'' as posts from 35 pre-selected high-quality news accounts on Twitter/X, while all others were labeled as ``non-news'' in our pipeline. We acknowledge that there might also exist news-related content among the non-news posts, for example, from local news organization accounts or individual journalists/editors. Because news posts are relatively rare in the control feed, we conduct a robustness check using engagement with all political posts as a broader outcome. Our goal is to test whether the observed treatment effects are driven by the low baseline frequency of specifically news posts in sample feeds prior to treatments.

We replicate the main GLMM analyses using political-post engagement as the dependent variable. The model specification follows the main analysis model. As shown in Table~\ref{tab:political_count} and Table~\ref{tab:political_rate}, the results are consistent with the main news-engagement analysis, suggesting that the treatment effects are not solely an artifact of the sparse baseline presence of news posts in the feed.

All four treatments remain statistically significant compare to the control condition ($p<0.001$) on both Count and Rate scales. The rank ordering of the four treatments varies slightly: the effect of Celebrity is now closer to Inject-2, rather than closer to Inject-4 as in the main analysis in Section \ref{sec:relative}. This difference likely reflects the fact that, for this robustness check, we did not rerun the full pipeline with manipulations applied specifically to political posts. The Celebrity effect is therefore shown as a conservative comparison. 

\begin{table*}[t]
  \centering
  \caption{Poisson GLMM of Political Engagement Count. The dependent variable is the political engagement count per user per step. Coefficients are on the log scale; standard errors in parentheses. Intervention effects are relative to Control; demographic effects are relative to the reference categories (Female for gender; Bachelor's degree for education). All models include a user-level random intercept. $^{***}p < 0.001$, $^{**}p < 0.01$, $^{*}p < 0.05$.}
  \label{tab:political_count}\vspace{-3mm}
  \scriptsize
  \setlength{\tabcolsep}{3pt}
  \begin{tabular}{lccccc}
  \toprule
   & \textbf{Model 1} & \textbf{Model 2} & \textbf{Model 3} & \textbf{Model 4} & \textbf{Model 5} \\
  \textbf{Variable} & \textbf{Total} & \textbf{Likes} & \textbf{Comments} & \textbf{Reposts} & \textbf{Follows} \\
  \midrule
  Intercept          & $-1.183^{***}$ (0.014) & $-2.050^{***}$ (0.023) & $-2.907^{***}$ (0.030) & $-2.737^{***}$ (0.029) & $-4.304^{***}$ (0.054) \\
  \addlinespace[0.5em]
  \multicolumn{6}{l}{\textit{Treatments (Ref: Control)}} \\
  Celebrity          & $0.658^{***}$ (0.015) & $0.479^{***}$ (0.023) & $0.647^{***}$ (0.030) & $0.867^{***}$ (0.028) & $1.217^{***}$ (0.049) \\
  Inject-2           & $0.819^{***}$ (0.014) & $0.822^{***}$ (0.023) & $1.052^{***}$ (0.029) & $0.649^{***}$ (0.029) & $0.028$ (0.054) \\
  Inject-4           & $1.124^{***}$ (0.014) & $1.137^{***}$ (0.022) & $1.326^{***}$ (0.029) & $1.011^{***}$ (0.028) & $0.250^{***}$ (0.053) \\
  Upranking          & $2.391^{***}$ (0.014) & $2.826^{***}$ (0.021) & $2.159^{***}$ (0.028) & $1.599^{***}$ (0.027) & $1.243^{***}$ (0.049) \\
  \addlinespace[0.5em]
  \multicolumn{6}{l}{\textit{Controls (Ref: Female; Bachelor's degree)}} \\
  Age (Centered)     & $0.002^{***}$ (0.000) & $0.002^{***}$ (0.001) & $0.004^{***}$ (0.001) & $0.002^{*}$ (0.001) & $-0.011^{***}$ (0.001) \\
  Gender: Male       & $-0.033^{***}$ (0.008) & $-0.036^{**}$ (0.013) & $-0.048^{**}$ (0.016) & $-0.038^{*}$ (0.016) & $0.031$ (0.031) \\
  Edu: Doctoral      & $0.081^{***}$ (0.015) & $0.066^{**}$ (0.024) & $0.144^{***}$ (0.029) & $0.041$ (0.030) & $-0.204^{***}$ (0.060) \\
  Edu: Master's      & $0.050^{***}$ (0.013) & $0.058^{**}$ (0.021) & $0.072^{**}$ (0.025) & $-0.037$ (0.026) & $0.029$ (0.049) \\
  Edu: Some college  & $-0.083^{***}$ (0.012) & $-0.066^{***}$ (0.019) & $-0.181^{***}$ (0.023) & $-0.116^{***}$ (0.024) & $0.164^{***}$ (0.044) \\
  Edu: Some HS       & $-0.089^{***}$ (0.012) & $-0.037$ (0.019) & $-0.264^{***}$ (0.024) & $-0.180^{***}$ (0.024) & $0.120^{**}$ (0.044) \\
  Edu: No formal     & $-0.073^{***}$ (0.014) & $0.026$ (0.023) & $-0.329^{***}$ (0.028) & $-0.242^{***}$ (0.028) & $0.047$ (0.053) \\
  \midrule
  Observations       & 276{,}589 & 276{,}589 & 276{,}589 & 276{,}589 & 276{,}589 \\
  Users              & 13{,}835 & 13{,}835 & 13{,}835 & 13{,}835 & 13{,}835 \\
  \bottomrule
  \end{tabular}
\end{table*}
\begin{table*}[!ht]
  \centering
  \caption{Poisson GLMM of Political Engagement Rate (with offset $\log(\text{total\_activity})$). The dependent variable is the political engagement share per user per step. Coefficients are on the log scale; standard errors in parentheses. Intervention effects are relative to Control; demographic effects are relative to the reference categories (Female for gender; Bachelor's degree for education). All models include a user-level random intercept. $^{***}p < 0.001$, $^{**}p < 0.01$, $^{*}p < 0.05$.}
  \label{tab:political_rate}\vspace{-3mm}
  \scriptsize
  \setlength{\tabcolsep}{3pt}
  \begin{tabular}{lccccc}
  \toprule
   & \textbf{Model 1} & \textbf{Model 2} & \textbf{Model 3} & \textbf{Model 4} & \textbf{Model 5} \\
  \textbf{Variable} & \textbf{Total} & \textbf{Likes} & \textbf{Comments} & \textbf{Reposts} & \textbf{Follows} \\
  \midrule
  Intercept          & $-2.070^{***}$ (0.010) & $-2.860^{***}$ (0.016) & $-3.873^{***}$ (0.032) & $-3.696^{***}$ (0.031) & $-5.309^{***}$ (0.056) \\
  \addlinespace[0.5em]
  \multicolumn{6}{l}{\textit{Treatments (Ref: Control)}} \\
  Celebrity          & $0.710^{***}$ (0.011) & $0.514^{***}$ (0.017) & $0.721^{***}$ (0.033) & $0.955^{***}$ (0.031) & $1.316^{***}$ (0.052) \\
  Inject-2           & $0.664^{***}$ (0.011) & $0.674^{***}$ (0.016) & $0.942^{***}$ (0.032) & $0.524^{***}$ (0.031) & $-0.100$ (0.057) \\
  Inject-4           & $0.945^{***}$ (0.010) & $0.954^{***}$ (0.016) & $1.196^{***}$ (0.031) & $0.870^{***}$ (0.031) & $0.100$ (0.056) \\
  Upranking          & $1.903^{***}$ (0.010) & $2.265^{***}$ (0.015) & $1.728^{***}$ (0.030) & $1.152^{***}$ (0.030) & $0.798^{***}$ (0.052) \\
  \addlinespace[0.5em]
  \multicolumn{6}{l}{\textit{Controls (Ref: Female; Bachelor's degree)}} \\
  Age (Centered)     & $0.001^{***}$ (0.000) & $0.002^{***}$ (0.000) & $0.004^{***}$ (0.001) & $0.001$ (0.001) & $-0.012^{***}$ (0.001) \\
  Gender: Male       & $-0.028^{***}$ (0.005) & $-0.023^{**}$ (0.008) & $-0.048^{**}$ (0.018) & $-0.040^{*}$ (0.018) & $0.036$ (0.032) \\
  Edu: Doctoral      & $0.064^{***}$ (0.009) & $0.048^{**}$ (0.015) & $0.142^{***}$ (0.032) & $0.045$ (0.033) & $-0.221^{***}$ (0.064) \\
  Edu: Master's      & $0.016^{*}$ (0.008) & $0.023$ (0.013) & $0.047$ (0.028) & $-0.061^{*}$ (0.028) & $0.007$ (0.052) \\
  Edu: Some college  & $-0.061^{***}$ (0.007) & $-0.043^{***}$ (0.012) & $-0.168^{***}$ (0.026) & $-0.099^{***}$ (0.026) & $0.186^{***}$ (0.047) \\
  Edu: Some HS       & $-0.068^{***}$ (0.007) & $-0.013$ (0.012) & $-0.246^{***}$ (0.026) & $-0.165^{***}$ (0.026) & $0.144^{**}$ (0.047) \\
  Edu: No formal     & $-0.085^{***}$ (0.009) & $0.017$ (0.014) & $-0.341^{***}$ (0.031) & $-0.250^{***}$ (0.031) & $0.047$ (0.056) \\
  \midrule
  Observations       & 273{,}134 & 273{,}134 & 273{,}134 & 273{,}134 & 273{,}134 \\
  Users              & 13{,}835 & 13{,}835 & 13{,}835 & 13{,}835 & 13{,}835 \\
  \bottomrule
  \end{tabular}
\end{table*}

\subsection{Action-Specific Effect Comparisons}
\label{app:action_specific}
For each engagement action type, we fit two complementary models (rate and count). For each fitted model, we compute the treatment-level estimated marginal means (EMMs) on the response scale, averaging over the sample distribution of gender, age, and education. The resulting EMMs are reported in Table \ref{tab:news_action_rate_count}. Notably, these tables contain model-based predictions, not raw per-treatment sample means. 

The total-engagement results show the same relative treatment effect ordering in both count and rate models. This ordering is stable for likes and comments. Reposts and follows show a more action-specific pattern: Celebrity exceeds Inject-4 in the count model and has the highest effect in the rate model. 

\begin{table*}[htbp]
\centering
\caption{News engagements by treatment and action type. Panel A reports the expected proportion of all engagements per step with news, expressed as a percentage (rate model with $\log(\text{total\_activity})$ offset). Panel B reports the expected absolute number of news-directed engagements (count model). Both are model-based EMMs on the response scale, averaged over gender and education levels with equal weights, with age at its sample mean. \textbf{Bold} values mark the largest effect within each action type column and within each treatment row.}
\label{tab:news_action_rate_count}

\begin{minipage}[t]{0.49\textwidth}
\centering
\textbf{A. News-Action Rate per Step by Treatment and Action Type}

\small
\setlength{\tabcolsep}{4pt}
\renewcommand{\arraystretch}{1.05}
\begin{tabular}{lrrrrr}
\toprule
Treatment & Like & Comment & Repost & Follow & Total \\
\midrule
Control    & 0.58\% & 0.12\% & 0.17\% & 0.05\% & \textbf{1.07\%} \\
Inject-2   & 3.61\% & 1.75\% & 1.21\% & 0.18\% & \textbf{7.62\%} \\
Celebrity  & 4.72\% & 2.46\% & \textbf{4.15\%} & \textbf{1.15\%} & \textbf{14.98}\% \\
Inject-4   & 5.86\% & 2.47\% & 2.07\% & 0.30\% & \textbf{12.05\%} \\
Upranking  & \textbf{26.83\%} & \textbf{5.13}\% & 3.41\% & 1.01\% & \textbf{40.81\%} \\
\bottomrule
\end{tabular}
\end{minipage}
\hfill
\begin{minipage}[t]{0.49\textwidth}
\centering
\textbf{B. News-Action Count per Step by Treatment and Action Type}

\small
\setlength{\tabcolsep}{4pt}
\renewcommand{\arraystretch}{1.05}
\begin{tabular}{lrrrrr}
\toprule
Treatment & Like & Comment & Repost & Follow & Total \\
\midrule
Control    & 0.013 & 0.003 & 0.005 & 0.001 & \textbf{0.027} \\
Inject-2   & 0.092 & 0.053 & 0.037 & 0.006 & \textbf{0.227} \\
Celebrity  & 0.098 & 0.061 & 0.100 & 0.029 & 0.347 \\
Inject-4   & 0.151 & 0.076 & 0.064 & 0.010 & \textbf{0.363} \\
Upranking  & \textbf{1.054} & \textbf{0.209} & \textbf{0.141} & \textbf{0.043} & \textbf{1.618} \\
\bottomrule
\end{tabular}
\end{minipage}

\end{table*}

\begin{figure}[t]
  \centering
  \includegraphics[width=\columnwidth]{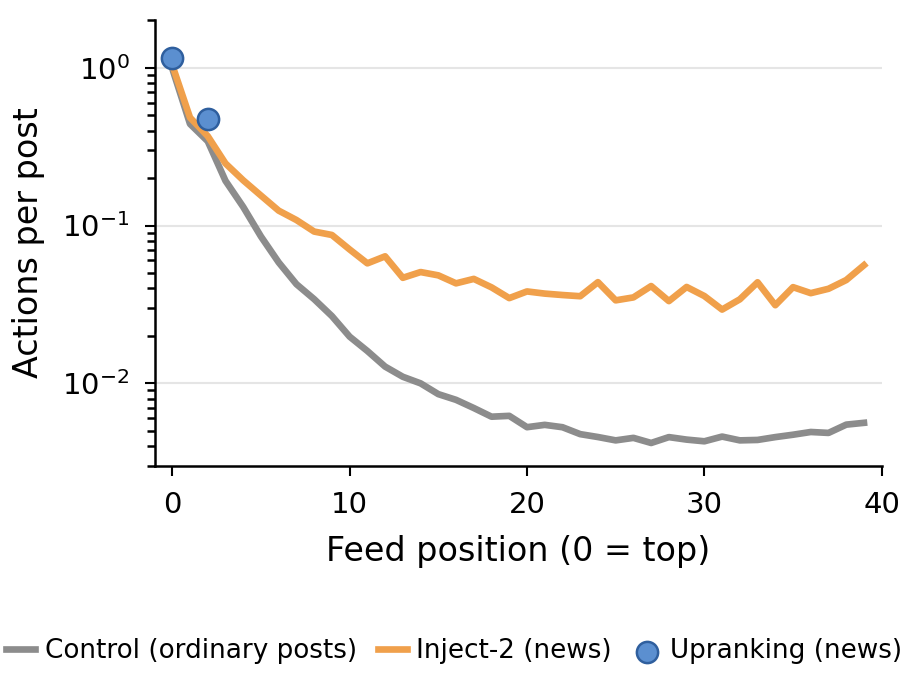}
  \caption{Engagement actions per post by feed position (log scale), for ordinary posts in Control (grey) and the two injected news posts in Inject-2 (orange), both of which occupy random positions, and for the same injected posts under Upranking at positions 0 and 2 (blue).}
  \label{fig:position}
\end{figure}
\subsection{Demographic Heterogeneity}
\label{app:demographic}
We investigate whether the observed treatment effects are universal or driven by specific user demographic attributes. The analysis uses the same data that drives the primary Poisson mixed-effects models in Section \ref{sec:overall}. Demographic variables enter the model as follows: Gender is a two-level factor (Female reference, Male). Education is a six-level factor. Age is mean-centered and treated as continuous so that the treatment main effect is interpretable at the sample mean age rather than at age zero. For each demographic axis, we fit a Poisson mixed-effects model. The base model is identical to the primary specification in Section \ref{sec:stats}. The user-level random intercept and Poisson link are unchanged. Each model adds one interaction term, as shown in Table~\ref{tab:interaction_models}.

\begin{table*}[ht]
\centering
\caption{Additional interaction terms and newly introduced parameters by model.}
\label{tab:interaction_models}
\begin{tabular}{lll}
\toprule
Model & Added term & New parameters \\
\midrule
\texttt{m\_gender}    & treatment $\times$ gender       & 4 \; (one per non-reference treatment) \\
\texttt{m\_education} & treatment $\times$ education    & 20 \; (4 treatments $\times$ 5 non-reference education levels) \\
\texttt{m\_age}       & treatment $\times$ age\_centered & 4 \\
\bottomrule
\end{tabular}
\end{table*}

First, we estimate the statistical significance of demographic attributes. Type III Wald chi-square tests confirm the significance of interaction terms for Gender ($\chi^{2} = 23.64$, $p < .001$, Age ($\chi^{2} = 22.24$, $p < .001$) and Education ($\chi^{2} = 72.79$, $p < .001$). It suggests that the magnitude of treatment effects varies slightly across user groups. However, these statistical significance largely reflects our large sample size ($N=276,589$ user-step observations; 13,835 active users). 

To assess the practical interaction effects, we examine the estimated coefficients. For gender, the largest interaction is Celebrity $\times$ Male ($\beta = -0.10$, $p = .10$), implying that the predicted Celebrity effect is about 10\% smaller for male users than for female users. For Age, all interaction coefficients are very close to zero. This means that each additional year of age changes the treatment effect by less than 0.5\%, which is substantively minor. Overall, although the interaction terms are significant, their magnitudes are small. It suggests that demographic differences slightly shift the treatment effect size, but do not substantially change the overall pattern. The complete statistical results are shown in Table~\ref{tab:interaction_summary}.

\begin{table}[ht]
\centering
\caption{Statistical and practical significance of treatment interactions.}
\label{tab:interaction_summary}
\small
\setlength{\tabcolsep}{4pt}

\begin{minipage}[t]{0.48\textwidth}
\centering
\textbf{A. Joint test statistics}

\begin{tabular}{lrr}
\toprule
Term & $\chi^2$ & $\chi^2 / df$ \\
\midrule
Treatment & 44{,}445.33 & 11{,}111.33 \\
Trt $\times$ Gender & 23.64 & 5.91 \\
Trt $\times$ Age & 22.24 & 5.56 \\
Trt $\times$ Educ. & 72.79 & 3.64 \\
\bottomrule
\end{tabular}
\end{minipage}
\hfill
\begin{minipage}[t]{0.48\textwidth}
\centering
\textbf{B. Interaction coefficients}

\begin{tabular}{lrr}
\toprule
Term & $\beta$ & Shift \\
\midrule
Celeb $\times$ Male & -0.101 & 0.904 \\
Inj-2 $\times$ Male & -0.062 & 0.940 \\
Inj-4 $\times$ Male & -0.008 & 0.992 \\
Up $\times$ Male    & +0.010 & 1.010 \\
Celeb $\times$ Age  & +0.003 & 1.003/year \\
Inj-2 $\times$ Age  & +0.005 & 1.005/year \\
Inj-4 $\times$ Age  & +0.005 & 1.005/year \\
Up $\times$ Age     & +0.001 & 1.001/year \\
\bottomrule
\end{tabular}
\end{minipage}

\end{table}

\begin{figure*}[!ht]
  \centering
  \includegraphics[width=\textwidth]{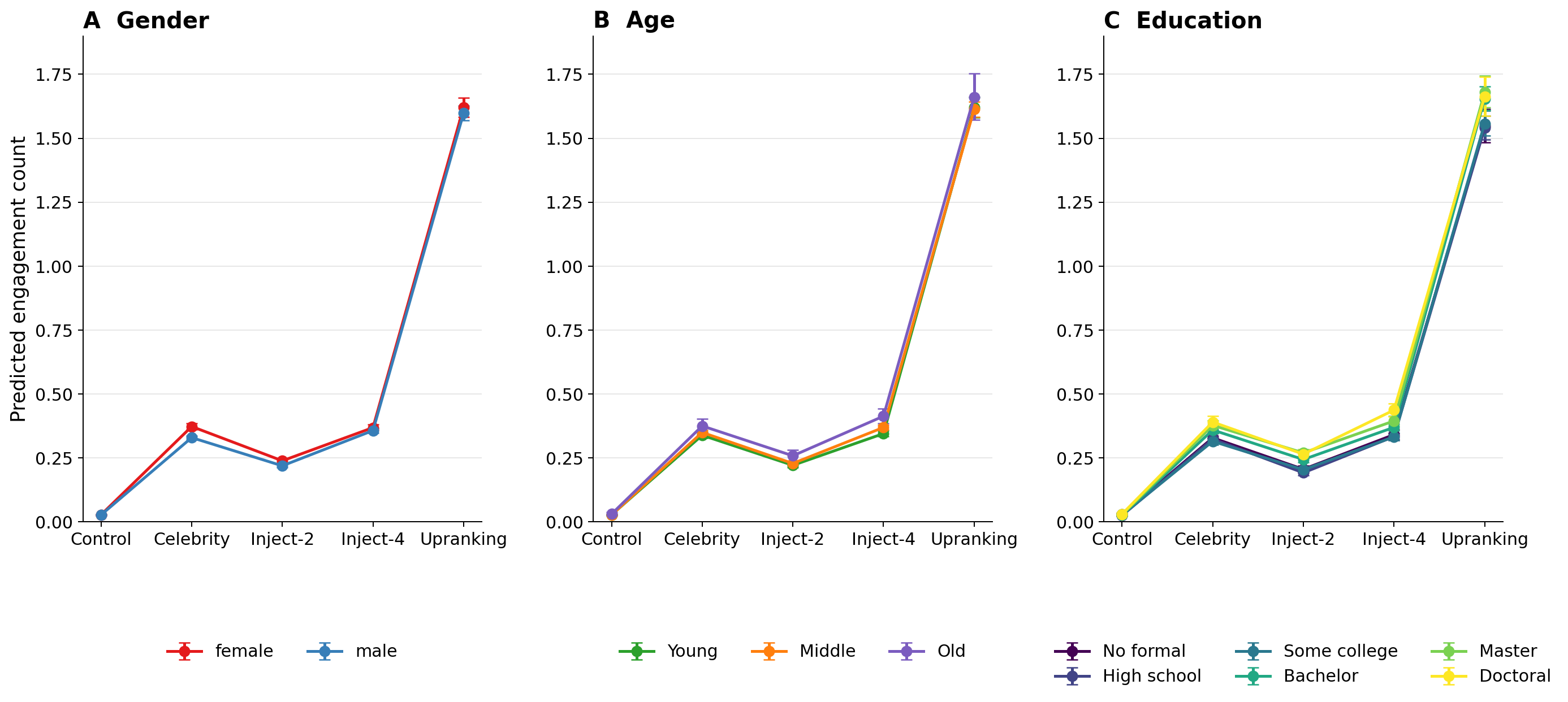}
  \caption{Interaction Analysis of Treatment Effects by Demographic Heterogeneity. The panels show two-way interactions between treatments and agent demographics: (A) Gender, (B) Age and (C) Education. Points represent the Estimated Marginal Means of engagement counts, with error bars indicating 95\% CI. The largely parallel patterns indicate the consistent effectiveness of the treatments across individual demographic differences.} 
  \label{fig:2ways_inter}
\end{figure*}

Second, we present estimated marginal means (EMMs) of news engagement under each treatment in Figure \ref{fig:2ways_inter} to further support the claim of negligible interaction effects. As shown by the largely parallel lines across all panels, the relative effectiveness of treatments are robust to demographic groups. For example, upranking is dominate across all age groups and education levels. Combined with the low chi-square values, these results confirm that the engagement shift patterns are universal across the entire user population. 

\begin{table}[t]
\centering
\small
\caption{Dose-normalized comparison with Yu et al. (2024). Dose is the change in news exposure, effect is the change in news share, and efficiency is the effect-to-dose ratio.}
\label{tab:validation-efficiency}
\begin{tabular*}{\columnwidth}{@{\extracolsep{\fill}}lccc@{}}
\hline
Metric & Yu et al. & Inject-2 & Inject-4 \\
\hline
Dose (pp)   & 11.99  & 4.93  & 9.86 \\
Effect (pp) & +16.29 & +6.55 & +10.98 \\
\hline
Efficiency  & 1.36   & 1.33  & 1.11 \\
\hline
\end{tabular*}
\end{table}

\subsection{Engagement by Feed Position}
\label{app:position}
Figure \ref{fig:position} shows the mean number of actions a post receives as a function of its feed position (position 0 is the top of a feed). Ordinary posts in Control and the two news posts injected under Inject-2 appear randomly in the feed. Thus, each curve is an engagement distribution by feed position. The blue points mark the same injected posts under Upranking at positions 0 and 2.

First, engagement drops sharply with position. In the control feed, agents engage with 85\% of ordinary posts at position 0 but with only 0.7\% of posts at position 10 or later, and 70\% of all actions in the control condition go to the first three positions. Second, ordinary and news posts follow a similar attention pattern, receiving 0.99 and 0.94 actions per post, respectively, at position 0. Third, at position 0, news in the Upranking condition receives 1.15 actions per post, compared with 1.04 for Inject-2 news that appears there by chance.

\section{Details of Validation}
\label{app:validation}
\paragraph{Model Specification and Estimand Matching}
In \citet{yu2024nudging}'s study, the primary treatment effect in the reference study is estimated using a user-level OLS model:
$$Y_{i, w23} \sim \beta_0 + \beta_1 \text{Treatment}_i + \gamma \mathbf{X}_i + \delta Y_{i, w1} + \epsilon_i$$
where $Y_{i, w23}$ is ratio-of-sums for user $i$ during the treatment period (week 2 and week 3; Day 8-21): total news videos watched / total videos watched × 100. The vector $\mathbf{X}_i$ includes demographics (age, gender, education, race, and political party), and $Y_{i, w1}$ is the baseline news share from the pre-treatment week.

\paragraph{Estimand Alignment.}
While our primary analysis uses a Poisson mixed-effects model to leverage the granularity of step-level data, we validated our findings by re-running our analysis using Yu et al.'s exact OLS specification (treating the entire simulation as a single aggregate period). We found that the resulting treatment effects ($+11.12$ pp) are nearly identical to our primary Poisson-offset estimates ($+10.98$ pp). This cross-model convergence demonstrates that our per-step engagement rate is a statistically valid proxy for the engagement rate measured in the reference human study.

\paragraph{Reconstructing the Intervention ``Dose''}
Yu et al. report a descriptive between-group difference in news recommendation share of approximately $8.73$ pp. However, to maintain internal consistency with the outcome metric (sum of news engagement divided by the sum of total engagement during their treatment period), we recalculated the ``dose'' using the same raw data and sample filtering used for their treatment effect estimation.

By computing the user-level ratio-of-sums for recommendations during the treatment window (Days 8-21), we found a mean news exposure of 18.30\% for the treatment group and 6.31\% for the control group, yielding a dose of 11.99 pp. In our analysis, we use an identical aggregation method for our simulation data, we measured a baseline news exposure of 1.29\%, which increased to 6.22\% in Inject-2 (dose: 4.93 pp) and 11.15\% in Inject-4 (dose: 9.86 pp).

\paragraph{Dose-Normalized Effect Calculation}
Accordingly, accounting for the differing doses, we compare the dose-normalized, defined as the percentage point increase in news share per 1 pp increase in news exposure (dose). The results show a remarkable convergence: the dose-normalized values for our simulated agents (1.11--1.33) are closely aligned with the human subject benchmark (1.36). See Table \ref{tab:validation-efficiency} for details.

\paragraph{Count-Based Effect Discussion}
Their absolute engagement counts are not directly comparable to ours on the magnitude level, because the estimation scale is different ($1.01$ in Yu's ``active day'' vs. $0.200$--$0.336$ in our ``single step''). Besides, our pipeline generate random feed at each step, while Yu’s field experiment involves a feedback loop where user interaction influences future recommendations, which may compound the treatment effect over time.

\end{document}